\documentclass[fleqn,usenatbib]{mnras}

\usepackage{newtxtext,newtxmath}

\usepackage{aas_macros}
\usepackage{orcidlink}
\usepackage{fontawesome}
\usepackage{float}

\usepackage[T1]{fontenc}

\DeclareRobustCommand{\VAN}[3]{#2}
\let\VANthebibliography\thebibliography
\def\thebibliography{\DeclareRobustCommand{\VAN}[3]{##3}\VANthebibliography}

\usepackage{graphicx}	
\usepackage{amsmath}	
\usepackage{amssymb}	
\usepackage{caption}
\usepackage{subcaption}
\usepackage{graphicx}
\usepackage{subcaption}
\usepackage{multirow}
\usepackage{array}
\usepackage{comment}
\usepackage{xcolor}
\usepackage{booktabs}
\newcolumntype{P}[1]{>{\centering\arraybackslash}p{#1}}

\usepackage{ulem}

\newcommand{\bn}{{\bf n}}
\newcommand{\bv}{{\bf v}}
\newcommand{\bnv}{{\bn\cdot\frac{\bv_0}{c}}}

\newcommand{\ds}{\displaystyle}

\newcommand{\rbr}[1]{\left(#1\right)}

\newcommand{\dd}{\mathrm d}
\newcommand{\mm}{m_{1,2}}
\newcommand{\calA}{{\mathcal{A}}}

\definecolor{Mgreen}{rgb}{0.1, 0.69,0.16}
\definecolor{MyOrange}{rgb}{0.97, 0.4, 0.29}

\newcommand{\be}{\begin{equation}}
\newcommand{\ee}{\end{equation}}
\newcommand{\bea}{\begin{eqnarray}}
\newcommand{\eea}{\end{eqnarray}}
\newcommand{\nn}{\nonumber}

\newcommand{\de}{\mathrm{d}}

\title[Cosmic dipole from combined GW estimators]{Combining chirp mass, luminosity distance and sky localisation from gravitational wave events to detect the cosmic dipole}

\author[Grimm]{N.~Grimm$^{1}$\orcidlink{0000-0001-9602-0599}, M.~Pijnenburg$^{1}$\orcidlink{0000-0002-1713-7412}, S.~Mastrogiovanni$^{2}$\orcidlink{0000-0003-1606-4183},
 C.~Bonvin$^{1}$\orcidlink{0000-0002-5318-4064}, S.~Foffa$^{1}$\orcidlink{0000-0002-4530-3051},
 G.~Cusin$^{1,3}$ \orcidlink{0000-0001-6046-1237}
\\
$^{1}$Universit\'e de Gen\`eve, D\'epartement de Physique Th\'eorique and Gravitational Wave Science Center, 24 quai Ernest-Ansermet, \\CH-1211 Gen\`eve 4, Switzerland\\
$^{2}$ INFN, Sezione di Roma, I-00185 Roma, Italy\\
$^{3}$Sorbonne Université, CNRS, UMR 7095, Institut d'Astrophysique de Paris, 75014 Paris, France \\
}
\date{Accepted XXX. Received YYY; in original form ZZZ\\
ET-0257A-23}
\pubyear{2023}

\begin{document}
\label{firstpage}
\pagerange{\pageref{firstpage}--\pageref{lastpage}}
\maketitle

\begin{abstract}
A key test of the isotropy of the Universe on large scales consists in comparing the dipole in the Cosmic Microwave Background (CMB) temperature with the dipole in the distribution of sources at low redshift. Current analyses find a dipole in the number counts of quasars and radio sources that is 2-5 times larger than expected from the CMB, leading to a tension reaching 5\,$\sigma$. In this paper, we derive a consistent framework to measure the dipole independently from gravitational wave (GW) detections. We exploit the fact that the observer velocity does not only change the distribution of events in the sky, but also the luminosity distance and redshifted chirp mass, that can be extracted from the GW waveform. We show that the estimator with higher signal-to-noise ratio is the dipole in the chirp mass measured from a population of binary neutron stars. Combining all estimators (accounting for their covariance) improves the detectability of the dipole by 30-50 percent compared to number counting of binary black holes alone. 
We find that a few $10^6$ events are necessary to detect a dipole consistent with the CMB one, whereas if the dipole is as large as predicted by radio sources, it will already be detectable with $10^5$ events, which would correspond to a single year of observation with next generation GW detectors. GW sources provide therefore a robust and independent way of testing the isotropy of the Universe.
\end{abstract}




\section{Introduction}

One of the basic assumptions of the $\Lambda$CDM cosmological model is that our Universe is homogeneous and isotropic on large scales. The latter follows from the high large-scale isotropy in observations of the Cosmic Microwave Background (CMB) temperature and of the large-scale structure of the Universe. Combining this with the cosmological principle further leads to the homogeneity of the Universe. In the $\Lambda$CDM model, the observed dipole anisotropy in the CMB temperature~\citep{Planck:2018nkj} is due to the fact that we, as observer, are moving with respect to the homogeneous and isotropic background.\footnote{Note that in the $\Lambda$CDM model, fluctuations in the matter density with respect to the homogeneous and isotropic background also generate a dipole anisotropy, but this contribution is negligible with respect to the one due to the kinematic dipole.} 

If this picture is correct, then the same motion should induce a dipole in the distribution of sources, due to aberration and magnification effects~\citep{Ellis1984}. This idea has been put to the test in the past years, through measurement of the dipole in the number counts of quasars and radio sources~\citep{Colin:2017juj,Bengaly_2018,Secrest:2020has,Siewert:2020krp,Secrest:2022uvx} at redshifts $0 \leq z \lesssim 3$. The direction of these dipoles is well aligned with that of the CMB, however the amplitude is 2-5 times larger than expected, leading to a tension with the CMB dipole reaching up to 5.1\,$\sigma$~\citep{Secrest:2022uvx}. {Supernovae type Ia light curves (see e.g.~\cite{Riess:1994nx}) provide an alternative means of assessing the dipole anisotropy~\citep{Bonvin:2006en}. Indeed, measurements from supernovae catalogs performed in~\cite{Singal:2021crs, Horstmann:2021jjg} have found a dipole again aligned with the CMB. However, they lead to an amplitude either compatible with the measurement from radio and quasar sources~\citep{Singal:2021crs}, or even lower than the CMB value~\citep{Horstmann:2021jjg}. Another measurement in~\cite{Sorrenti:2022zat} shows an amplitude consistent with the CMB, but a strong tension in the direction. Hence, the inconclusive measurements of the dipole from type Ia supernovae do not resolve any tensions.}

The discrepancy in the dipole amplitude could be due to systematic effects in the quasar and radio source data sets, to an imperfect theoretical modelling of the expected dipole in the number counts, which currently neglects evolution effects~\citep{Dalang:2021ruy,Guandalin:2022tyl}, or to a violation of isotropy in our Universe. In this last case, the large dipole in the quasars and radio sources would not be solely due to the observer velocity, but it would have an intrinsic part generated by a large local anisotropy in the Universe (inconsistent with the $\Lambda$CDM predictions).  

One way to test these scenarios is to use other data sets to measure the dipole at low redshift, and see if the result is consistent with the CMB dipole, or with the dipoles from quasars and radio sources. A promising avenue is to use gravitational waves (GWs) from binary systems of black holes (BBH) or neutron stars (BNS). In~\citet{PhysRevD.97.103005}, it was proposed to look for a dipole in the luminosity distance measured from GWs. This paper does not model the kinematic contribution to the luminosity distance dipole, but it forecasts how large a dipole should be to be detected (independently of its origin). In~\cite{Mastrogiovanni:2022nya}, some of us proposed to use the distribution of BBH to measure the cosmic dipole. We derived a modelling of the signal, taking into account aberration and threshold effects, and showed that with the next generation of interferometers (XG), like the Einstein Telescope (ET) and the Cosmic Explorer (CE) it will be possible to confidently detect the dipole. In~\citet{Kashyap:2022ibx}, the authors propose to use both the number counts of GWs and the chirp mass to measure the dipole. They apply their method to current data and show that no dipole anisotropy is detected. This is in agreement with~\cite{Stiskalek:2020wbj,Essick:2022slj, Payne:2020pmc}, who found no evidence for an anisotropy in the distribution of current GW sources.

In this paper, we derive a consistent framework to use GW detections to optimally measure the cosmic dipole. In particular, we exploit three of the quantities that can be extracted from the GW waveform and amplitude: the angular position of the binary system, its luminosity distance, and its redshifted chirp mass. These three quantities are all affected by the observer velocity and can therefore be combined to measure the velocity in an optimal way. We build on the results of~\cite{Mastrogiovanni:2022nya} to derive a modelling of the mean luminosity distance and the mean chirp mass per angular pixel. This modelling accounts not only for aberration, but also for threshold effects. We show that the distance and chirp mass dipoles are correlated with each other, but both are independent of the number count dipole. Combining them  therefore requires taking into account this correlation, in order not to overestimate the constraints. We also show that threshold effects can be robustly modelled and computed, which is essential in order not to bias the measurement of the observer velocity.

We apply our framework to synthetic catalogues of binary systems (BBH and BNS) that would be observed by the next generation of interferometers, like ET and CE, and we forecast how well the dipole can be detected in these catalogues. We explore two scenarios: one where the dipole would be purely kinematic and therefore consistent with the CMB value, and a second one where the dipole is consistent with the results from radio sources. For this second scenario, we take the extreme case of an observer velocity that would be 5 times larger than the CMB one~\citep{Bengaly_2018}, and we call this the ``AGN case''. Note that if we take the AGN case at face value, the dipole cannot be only due to the observer velocity (that would be inconsistent with the one extracted from the CMB), but it would have a large intrinsic part due to a strong anisotropy in the large-scale structure, as discussed above. However, when assessing the detectability of such a dipole, it does not matter if it is purely of kinematic origin, or if there is also an intrinsic contribution. Therefore we can simulate the AGN dipole as if it were due to an observer velocity which is 5 times larger than expected. 

We find that the chirp mass from BNSs provides the estimator with lower variance, i.e.\ larger signal-to-noise ratio, followed by the number counts of BBHs and BNSs. This is not surprising given the predominant impact of this parameter on the BNS inspiral waveform; the possibility of exploiting this fact has already been subject of various investigations~\citep{Finke:2021eio,Chernoff:1993th,Taylor:2011fs,Taylor:2012db}. Our analysis further shows that, combining all estimators, we could detect a dipole consistent with the CMB one at $> 1\,\sigma$ significance with more than $10^6$ events, achieved with 5 years of observations of ET and CE. If the dipole is as large as predicted in the AGN case, then we can detect it at $> 3\,\sigma$ significance with $10^5$ events already, achieved in a year of observation.

The rest of the paper is structured as follow: in Sec.~\ref{sec:model} we derive a theoretical modelling of the luminosity distance and chirp mass dipole. In Sec.~\ref{sec:estimators} we define estimators of these dipoles and we compute their variance and covariance. In Sec.~\ref{sec:sim} we measure the dipoles from our synthetic catalogues of events and assess the detectability of the six estimators (three for BBHs and three for BNSs) and their combination. In Sec.~\ref{sec:Fisher} we forecast how well the observer velocity can be measured with XG detectors and we show that an imperfect knowledge of threshold effects does not degrade the constraints significantly. We conclude in Sec.~\ref{sec:conclusion}.

\section{Theoretical modelling of the luminosity distance and chirp mass dipole}
\label{sec:model}

To compute the impact of the observer velocity on the luminosity distance and the chirp mass measured from GWs, we follow the same steps as in~\cite{Mastrogiovanni:2022nya}, which computed the dipole in the number of events. Since the observer velocity is significantly smaller than the speed of light, we keep only terms at linear order in $v_0/c$ in the derivation.

\subsection{Luminosity distance }

The luminosity distance of a source situated at conformal distance $r$ from the observer is defined as the ratio between the intrinsic luminosity of the source, $L$, and the observed flux, $F$
\begin{align}
 d_L(r,\bn)=\sqrt{\frac{L}{4\pi F(r,\bn)}}\, ,   
\end{align}
where $\bn$ denotes the direction in which the observer sees the source. The observer velocity impacts the flux measured by the observer and consequently modifies the measured luminosity distance. At first order in $v_0/c$ the luminosity distance is given by (see~\cite{Bonvin:2005ps}~\footnote{Note that here we are interested in the luminosity distance at fixed conformal distance $r$, which is given by~Eq.~(55) of~\cite{Bonvin:2005ps}. Eq.~(59) gives instead the luminosity distance at fixed observed redshift. Note also a sign difference due to the fact that in~\cite{Bonvin:2005ps} $\bn$ denotes the direction of propagation of the signal, whereas here $\bn$ is the direction of observation, which points in the opposite direction.})
\begin{equation}
    d_L(r, \bn) = \bar{d}_L(r) \ \bigg(1-\bnv\bigg)
    \label{dLv0} \,,
\end{equation}
where $\bar{d}_L(r)$ denotes the background luminosity distance in a homogeneous and isotropic universe. Note that other perturbations, in particular the peculiar velocity of the source, affect the luminosity distance besides the observer velocity, see~\cite{Bonvin:2005ps} for the full expression. However, here we are interested in the dipole which is strongly dominated by the observer velocity, and other contributions are negligible compared to typical uncertainties of about 20\% for luminosity distance measurements from GW events~\citep{2020ApJ...896L..32C,2022arXiv220702771I}. 

The luminosity distance enters directly in the amplitude of the GWs, which decays as $1/d_L$. Combining measurements from different interferometers allows one to measure $d_L$ for each binary system.  We can then compute the mean luminosity distance from a population of binary systems in direction $\bn$.
\begin{equation}
\ds
d^{\rm mean}_L (\bn)\equiv \frac{\int {\rm d}r\ d_L(r,\bn)\ \frac{{\rm d}N_{\rm det}}{{\rm d}\Omega {\rm d}r}(r, \bn)}{\int {\rm d}r\ \frac{{\rm d} N_{\rm det}}{{\rm d}\Omega {\rm d}r}(r, \bn)}\,.\label{alternativedL}
\end{equation}
Here, 
\begin{align}
\label{eq:Ndetdef}
\frac{{\rm d}N_{\rm det}}{{\rm d}\Omega {\rm d}r}(r, \bn)=\int {\rm d}m_{1,2}\frac{{\rm d}N_{\rm det}}{{\rm d}\Omega {\rm d}r {\rm d}m_{1,2}}(r, \bn, m_{1,2},\rho>\rho_*)
\end{align}
denotes the number of events detected at distance $r$ with a signal-to-noise ratio (SNR) $\rho$ above a threshold value $\rho_*$, summed over all intrinsic binary masses $m_1$ and $m_2$~\footnote{Note that here we have dropped the dependence on the SNR threshold, $\rho_*$, in the left-hand side of Eq.~\eqref{eq:Ndetdef} for simplicity. When we refer to ``detected'' events, by definition we refer to events above the SNR threshold. We state the dependence explicitly only when it is not integrated over masses.}.
Following \cite{Mastrogiovanni:2022nya}, we adopt a simplified, zero Post-Newtonian (0PN) order for the SNR, which can be computed as~\citep{Finn:1992xs}
\be
\rho^2(r,\bn, m, \mathcal{M})= \frac{5}{96\pi^{4/3}} \frac{\Theta^2}{d_L^2(r,\bn)} (G \mathcal{M})^{5/3} \mathcal{F}\left(f^z_{\rm ISCO}(m)\right)\,,
\label{eq:SNR1}
\ee
where $\mathcal{M}$ is the redshifted chirp mass as measured in the detector frame (in the following we drop the ``redshifted'' for simplicity). In Eq.~\eqref{eq:SNR1}, $m$ is the total detector frame mass of the system and 
\be
f_{\text{ISCO}}\equiv\frac{1}{6\sqrt{6}(2\pi)}\frac{c^3}{Gm}\simeq  2.2. \text{kHz}\left(\frac{M_{\odot}}{m}\right)\,
\label{eq:isco}
\ee
is the GW frequency corresponding to the innermost stable circular orbit. As to $\Theta^2$, it is a geometrical factor that accounts for the binary inclination angle and average detector's antenna patterns:
\begin{equation}
    \Theta^2 = 0.2 \left[ \left(\frac{1+\cos^2 \iota}{2} \right)^2+\cos^2 \iota\right],
\end{equation}
where $\iota$ is the angle formed by the normal vector of the orbital plane and the line of sight. The numerical prefactor in the above equation represents an average value of all the detector's antenna patterns.
The function $\mathcal{F}$ is calculated from the integral of the power spectral density  $S_n(f)$ following~\cite{Maggiore:1900zz}.

Since the SNR~\eqref{eq:SNR1} depends on the luminosity distance and on the chirp mass, that are both affected by the observer velocity, the number of detected events above threshold $\rho_*$ will be modified by the observer velocity. These threshold effects add to the effect of aberration, which modifies the number of events per solid angle. As shown in~\cite{Mastrogiovanni:2022nya}, the final result at linear order in $v_0/c$ is given by
\begin{multline}
    	\frac{\mathrm d N_{\rm det}}{\mathrm d\Omega \mathrm dr\mathrm dm_{1,2}}(r, \bn, m_{1,2},\rho>\rho_*)   \\
     = \  \frac{\mathrm d \bar{N}_{\rm det}}{\mathrm d\Omega \mathrm dr \mathrm dm_{1,2}} \big(r, \mm, \rho_*\big)
	\Bigg\{1+\left[2+s\left(\frac{1}{3}+ {\calA}\right)\right]\bn\cdot\frac{\bv_0}{c}\Bigg\}\,,
 \label{DipoleCountDensity}
\end{multline}
with\footnote{Note that these functions depend on $m_{1, 2}$, while in \cite{Mastrogiovanni:2022nya}, $s$ and $\calA$ refer to mass integrated quantities}
\begin{align}
&s(r, \rho_*,m_{1,2})\equiv - \frac{\partial}{2\partial \ln\rho_*}\ln\rbr{\frac{\mathrm dN_{\rm det}\rbr{r,\rho>\rho_\ast,m_{1,2}}}{\mathrm d\Omega\,\mathrm  d r\,\mathrm dm_{1,2}}} \,,\label{newdefhats}\\
&{\calA}(r,\mm)\equiv \frac{1}{\mathcal F\rbr{\frac{f_{\rm ISCO}}{(1+z)}}}\rbr{\frac{2 f_{\rm ISCO}}{(1+z)}}^{-7/3}S_n\rbr{\frac{2 f_{\rm ISCO}}{(1+z)}}^{-1}\frac{2 f_{\rm ISCO}}{(1+z)}\,. \label{newdefhatA}
 \end{align}

Inserting Eqs.~\eqref{DipoleCountDensity} and~\eqref{dLv0} into Eq.~\eqref{alternativedL} and keeping only linear terms in $v_0/c$ we finally obtain for the mean luminosity distance
\begin{align}
d^{\rm mean}_L(\bn)=d_L^{(0)}\Big(1+{\cal D}d_L(\bn)\Big)\, ,    
\end{align}
where the monopole is given by
\begin{equation}
d^{(0)}_L\equiv\frac1{4\pi}\int\rm d\Omega\ d^{\rm mean}_L(\bn)\,.
    \label{dLMonopole}
\end{equation}
The fractional dipole is given by
\begin{equation}
    {\cal D}d_L(\bn)= - \alpha_{d_L}\left(\bn\cdot\frac{\bv_0}c\right) \,,  \label{Ddlr}
\end{equation}
where the prefactor $\alpha_{d_L}$ takes the form
\begin{equation}
    \alpha_{d_L} \equiv 1 - \int\!\mathrm dr\, \mathrm dm_{1,2}
    \ s\rbr{\frac 13 +\mathcal A}\left[
    \frac{\bar{d}_L(r)}{d_L^{(0)}}-1\right]p(r, \mm, \rho_*)\,, \label{defalphad}
\end{equation}
with $p(r, m_{1, 2}, \rho_*)$ being the mean probability density function of detected sources, 
\begin{align}
 p(r, m_{1, 2}, \rho_*)=\frac{\frac{{\rm d}\bar{N}_{\rm det}}{{\rm d}\Omega {\rm d}r {\rm d}m_{1, 2}}(r, m_{1, 2}, \rho>\rho_*)}{\int {\rm d}r\, {\rm d}m_{1, 2} \frac{{\rm d}\bar{N}_{\rm det}}{{\rm d}\Omega {\rm d}r {\rm d}m_{1, 2}}(r, m_{1, 2}, \rho>\rho_*)}\, . \label{ProbaDensity}
\end{align}
Note that in Eq.~\eqref{Ddlr} we have introduced the parameter $\alpha_{d_L}$ factorising out a minus sign in such a way that $\alpha_{d_L}=1$ in the case where there are no threshold effects, i.e.\ when $s=0$.

We see from Eq.~\eqref{defalphad} that if there are threshold effects but $s$ and $\mathcal{A}$ are constant in $r$, then the integral vanishes and $\alpha_{d_L}=1$. We can indeed rewrite the monopole as
\begin{align}
d^{(0)}_L&=\frac1{4\pi}\int\!\!\dd\Omega \left[\frac{\int\!\! \dd r {\rm }\,\dd m_{1,2}d_L(r,\bn)\frac{\dd N_{\rm det}}{\dd\Omega\dd r\dd m_{1,2}}(r,\bn,m_{1,2},\rho>\rho_*)}{\int\!\! \dd r {\rm }\,\dd m_{1,2}\frac{\dd N_{\rm det}}{\dd\Omega\dd r\dd m_{1,2}}(r,\bn,m_{1,2},\rho>\rho_*)} \right] \nonumber\\  
&=\int \dd r\,\dd m_{1,2}\bar{d}_L(r)p(r,m_{1,2},\rho_*)\, ,
\end{align}
since the dipole contributions in the luminosity distance and in the number of detected events vanish when integrated over direction (note that here we neglect terms or order $(v_0/c)^2$). Inserting this into Eq.~\eqref{defalphad} we see that the integral exactly vanishes if $s$ and $\mathcal{A}$ are independent on $r$.
Since these functions are expected to evolve slowly with $r$, we expect threshold effects to be partially suppressed, meaning that $\alpha_{d_L}$ will be close to 1, even in the case where $s$ is non-zero. In Sec.~\ref{sec:alpha}, we compute $\alpha_{d_L}$ from our population model of BNSs (where threshold effects are present), and compare this with measurements of $\alpha_{d_L}$ from our synthetic catalogues. We find that $\alpha_{d_L}$ can be well predicted and, as expected, that it is very close to 1. This is important, because $\alpha_{d_L}$ is fully degenerated with the observer velocity $v_0/c$ and if we do not know it, we cannot extract information on $v_0/c$ from a population of sources with threshold effects.

\subsection{Chirp mass}

The chirp mass that can be extracted from the GW is the redshifted chirp mass, i.e.\ the product of the intrinsic chirp mass of the binary system and the redshift. Since the latter is affected by the observer velocity, to first order, the redshifted chirp mass is given by 
\begin{equation}
    {\cal M}(r, \bn, m_{1, 2}) = {\cal \bar{M}}(r, m_{1, 2}) \bigg(1-\bnv\bigg)\,,
    \label{Mv0}
\end{equation}
where ${\cal \bar{M}}(r, m_{1, 2})$ is the background (redshifted) chirp mass in a homogeneous and isotropic universe. Note that, as for the luminosity distance, other perturbations contribute to the redshifted chirp mass, that can be neglected since they have a negligible impact on the dipole.
Analogously to Eq. \eqref{alternativedL}, we can compute the mean detected chirp mass in direction $\bn$, averaged over all binary systems (i.e.\ over all masses $m_{1,2}$) as 
\begin{align}
&\mathcal{M}^{\rm mean} (\bn)\equiv \nn \\
&\frac{\int \dd r\, \dd m_{1,2}\, \mathcal{M}(r,\bn,m_{1,2})\ \frac{\dd N_{\rm det}}{\dd\Omega \dd r \dd m_{1,2}}(r, \bn, m_{1,2})}{\int \dd r\ \frac{\dd N_{\rm det}}{\dd\Omega \dd r}(r, \bn)}\,. \label{dipoleM}  \end{align}
This can be expanded in powers of $\bnv$ as a monopole, $\mathcal{M}^{(0)}$, and dipole term: 
\begin{equation}
    \mathcal{M}^{\rm mean}(\bn) =  \mathcal{M}^{(0)}\Big(1+  {\cal D}\mathcal{M}(\bn)\Big)\,,
\end{equation}
where
\begin{equation}
 \mathcal{M}^{(0)} = \frac1{4\pi}\int\rm d\Omega\ \mathcal{M}^{\rm mean} (\bn)\,.
\end{equation}
Combining Eqs.~\eqref{DipoleCountDensity} and \eqref{Mv0} allows to write the fractional  dipole term as
\begin{equation}
        {\cal D}\mathcal{M}(\bn)= -\ \alpha_{\cal M}\ \left(\bn\cdot\frac{\bv_0}c\right) \,,
\end{equation}
where we introduced $\alpha_{\cal M}$ as the chirp mass analogue of $\alpha_{d_L}$:
\begin{equation}
    \alpha_{\cal M} \equiv 1 - \int\!\mathrm dr\, \mathrm dm_{1,2}
    \ s\rbr{\frac 13 +\mathcal A}\left[
    \frac{\bar{\cal{M}}(r, \mm)}{ \mathcal{M}^{(0)}}-1\right]p(r, \mm, \rho_*)\,.
    \label{defalphaM}
\end{equation}
As for the luminosity distance, we see that threshold effects are non-zero only if $s$ and $\cal{A}$ vary with distance $r$.

\cite{Kashyap:2022ibx} have also computed the dipole in the mean chirp mass (that they call mass intensity). Their modelling differs  from ours, since they assume that all events above a given mass threshold are detected. In practice, this is however not the case. What determines if an event is detected or not is the SNR, which depends not only on the mass of the binary system, but also on its distance from the observer, as can be seen from Eq.~\eqref{eq:SNR1}. This has an impact on the modelling of threshold effects, leading to a different expression for the dipole signal.

\subsection{Number counts}

The dipole in the GWs number count has been derived in~\citet{Mastrogiovanni:2022nya} and follows directly from Eq.~\eqref{DipoleCountDensity}. The total number of GW events detected in direction $\bn$ is given by 
\begin{align}
 N_{\rm det}(\bn)&\equiv \int \mathrm dr\,\mathrm dm_{1,2} \frac{\mathrm dN_{\rm det}}{\mathrm d\Omega\mathrm dr\mathrm dm_{1,2}}(r, \bn, m_{1,2})\nonumber\\
 &=N_{\rm det}^{(0)}\Big(1+  {\cal D}N_{\rm det}(\bn)\Big)\,,
 \label{CountDipole}
\end{align}
where
\begin{align}
   {\cal D} N_{\rm det}(\bn) = 2\, \alpha_{N}\ \bnv\,, \label{CountDipolealphaN}
\end{align}
with 

\begin{align}
  \alpha_{N}\equiv 1+\frac{1}{2}\int {\rm d} r \, {\rm d} m_{1, 2}\, s \bigg(\frac{1}{3}+{\cal A}\bigg)p(r, m_{1, 2}, \rho_*) \, .\label{defalphaN}
\end{align}

\section{Statistical estimators of the dipoles}
\label{sec:estimators}

We now build statistical estimators for the three dipole signals defined in Sec.~\ref{sec:model}. 

\subsection{Luminosity distance}
\label{sec:estimator_distance}

For each direction $\bn'$ in the sky, we can build the following observable
\begin{equation}\label{dlObservable}
    v_{d_L-\bn'}\equiv -3 \int \frac{{\rm d}\Omega}{4\pi }\ \left(\bn\cdot\bn'\right) {\cal D}d_L(\bn)\simeq \alpha_{d_L} \frac{v_0}c \cos{\theta'}\,,
\end{equation}
where $\theta'$ is the angle between the (a priori unknown) dipole velocity direction and $\bn'$.
This observable is maximized when evaluated along the dipole direction and, in the absence of threshold effects, it is exactly equal to the observer velocity, $v_0/c$, at the maximum.

Let us now build a statistical estimator for the observable $v_{d_L-\bn'}$. We divide the sky in $N_{\rm sky}$ angular pixels of same solid angle, and associate a vector $\bn_i$ pointing to the center of each pixel $i$. Within any pixel $i$, there are $N^i_{\rm det}$ detected events (labeled by $j$), whose corresponding luminosity distance is labeled by $(d_L)^i_j$. The estimator is then defined as
\begin{equation}\label{eq:dLstat}
    \hat{v}_{d_L-\bn'}\equiv \frac{-3}{\hat{d}_L N_{\rm sky} }\sum_{i=1}^{N_{\rm sky}}\sum_{ j=1}^{N^i_{\rm det}} \frac{(d_L)^i_j}{N^i_{\rm det}} (\bn_i\cdot\bn')\,,
\end{equation}
with
\begin{align}
\label{eq:dLhat}
\hat{d}_L\equiv \frac{1}{N_{\rm sky}}\sum_{i=1}^{N_{\rm sky}}\sum_{j=1}^{N^i_{\rm det}}\frac{(d_L)^i_j}{N^i_{\rm det}}\,. 
\end{align}
As shown in Appendix~\ref{app:variance}, if shot noise and the uncertainty in the measurement of the luminosity distance are subdominant (i.e.\ smaller than their mean), the estimator~\eqref{eq:dLstat} is unbiased, i.e.\ $\langle\hat{v}_{d_L-\bn'}\rangle=v_{d_L-\bn'}$. 

We then compute the variance of this estimator. It can be written as ratio of two variables $X/Y$ with
\begin{align}
X  = \frac{-3}{ N_{\rm sky} }\sum_{i=1}^{N_{\rm sky}}\sum_{ j=1}^{N^i_{\rm det}} \frac{(d_L)^i_j}{N^i_{\rm det}}  (\bn_i\cdot\bn')\,,
\label{eq:X}
\end{align}
and $Y=\hat{d}_L$ defined in Eq.~\eqref{eq:dLhat}. The variance of such a ratio can be written as (if the variances of $X$ and $Y$ are significantly smaller than their mean)
\begin{equation}
\label{eq:ratiovar}
{\rm var}\left(\frac{X}{Y}\right)\simeq\frac{\langle X\rangle^2}{\langle Y\rangle^2}\left[\frac{{\rm var}(X)}{\langle X\rangle^2}+\frac{{\rm var}(Y)}{\langle Y\rangle^2}-2\frac{{\rm cov}(X,Y)}{\langle X\rangle\langle Y\rangle}\right]\,.
\end{equation}
We can show that the third term in Eq.~\eqref{eq:ratiovar} is vanishing and the second one is smaller than the first one by a factor $({v_0}/{c})^2$ and can therefore be neglected (see Appendix~\ref{app:variance} for a detailed derivation). The variance of~\eqref{eq:dLstat} is therefore directly proportional to the variance of $X$ and can be written as
\begin{align}
&{\rm var}\left(\hat{v}_{d_L-\bn'}\right)=\langle \hat{v}^2_{d_L-\bn'}\rangle-\langle \hat{v}_{d_L-\bn'}\rangle^2\nn \label{eq:variancedL}\\
&=\frac{9}{\left(d^{(0)}_L\right)^2 N_{\rm sky}^2}\sum_{i=1}^{N_{\rm sky}}\sum_{a=1}^{N_{\rm sky}}\sum_{j=1}^{N^i_{\rm det}}\sum_{b=1}^{N^a_{\rm det}}
(\bn_i\cdot\bn')(\bn_a\cdot\bn')\nn\\
&\qquad\times\left(\left\langle\frac{(d_L)^i_j(d_L)^a_b}{N_{\rm det}^i N_{\rm det}^a} \right\rangle-
\left\langle\frac{(d_L)^i_j}{N_{\rm det}^i} \right\rangle\left\langle\frac{(d_L)^a_b}{N_{\rm det}^a} \right\rangle\right)\,.
\end{align}

\begin{figure*}
    \centering
    \includegraphics[width=\textwidth]{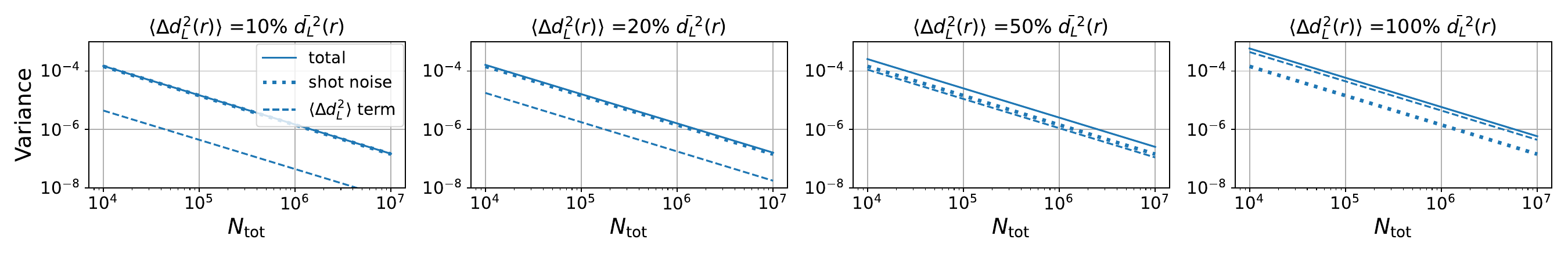}
    \caption{Variance of the luminosity distance estimator for BBHs, plotted as a function of the total number of events $N_{\rm tot}$. The different panels are for different values of the relative error on $d_L$ (assumed to be independent of redshift). We show separately the shot noise contribution and the distance-uncertainty contribution, as well as the total.}
    \label{fig:vardL_plots}
\end{figure*}

To compute this, we divide each solid angle $i$ into $N_r$ bins in distance $r$ of size $\Delta r$, and we rewrite the sum over $j$ as a sum over the $r$-bins. The sky is therefore divided into pixels of angular size $\Delta\Omega$ and radial size $\Delta r$. We denote by $N_p^{ik}$ the number of events detected in the pixel centered in direction $\bn_i$ and distance $r_k$ and by $(d_L)^i_k$ the corresponding luminosity distance. Due to shot noise, the number of events fluctuates around the mean from pixel to pixel,
\begin{align}
N_p^{ik}&=\frac{\dd N_{\rm det}(r_k,\bn_i)}{\dd\Omega\dd r}\Delta r\Delta\Omega=\bar{N}_p^k+\Delta N_p^{ik}\, , \label{eq:Npixel}\\
N^i_{\rm det}&=\bar{N}_{\rm det}+\sum_j \Delta N_p^{ij}\, .\label{eq:Nidet}
\end{align}
Note that here we neglect the fluctuations in $N_p^{ik}$ due to the uncertainty on sky localisation. We assume that this uncertainty is of $\sim 3$ degrees, significantly smaller than the size of the angular pixels that we consider.
In addition to shot noise, the variance of the estimator is affected by uncertainty in the measurement of the luminosity distance, which we write as
\begin{align}
(d_L)^i_k&=(\bar{d}_L)_k +(\Delta d_L)^i_k \, , \label{eq:dLpixel}
\end{align}
where $(\Delta d_L)^i_k$ is the error in the measurement of $d_L$ from one binary system, in angular bin $i$ and radial bin $k$. Note that in Eqs.~\eqref{eq:Npixel}-\eqref{eq:dLpixel} we neglect the contributions from the observer velocity that would lead to subdominant contributions to the variance. The main contribution comes indeed from a ``fluctuation'' of the monopole (due to shot noise and measurement uncertainty) that would mimic a dipole. Using the fact that errors on count and on luminosity distance are uncorrelated and that number counts follow a Poissonian statistics, we obtain for the variance (see Appendix~\ref{app:variance} for more detail) 
\begin{align}
{\rm var} \left(\hat{v}^2_{d_L-\bn'}\right)=&\frac{3 N_{\rm sky}}{\left(d^{(0)}_L\right)^2 N_{\rm tot}^2}\Bigg\{\sum_{k=1}^{N_r} \langle(\Delta d_L)^2_k \rangle\bar{N}_p^k \nn \\
&+\sum_{k=1}^{N_r}(\bar{d}_L)_k^2 \bar{N}_p^k-\frac{1}{\bar{N}_{\rm det}}\left(\sum_{k=1}^{N_r}(\bar{d}_L)_k\bar{N}_p^k\right)^2 \Bigg\}\,, \label{eq:var_final_discrete}
\end{align}
where $N_{\rm tot}$ is the total number of events detected and $\langle (\Delta d_L)_k\rangle$ denotes the typical error on luminosity distance measurement at distance $r_k$.

The term in the first line of Eq.~\eqref{eq:var_final_discrete} is the contribution from luminosity distance uncertainty to the variance of the estimator. If $\langle(\Delta d_L)_k\rangle$ is independent on distance, we see that the variance from this term scales as $1/N_{\rm tot}$, as expected for $N_{\rm tot}$ measurement of uncorrelated quantities. The contributions in the second line are due to shot noise: the mean distance in a given solid angle depends on the radial distribution of events. Since shot noise generates fluctuations in the number of events, a given solid angle can have more events closer to the observer, whereas another solid angle can have more events further away. This generates fluctuations in the mean distance, that can mimic the presence of a dipole. Note that since we are measuring the mean distance per angular pixel and not the sum of distances per pixel, we are not sensitive to the impact of shot noise on the total number of events per pixel, but rather to its impact on the radial distribution of sources. 

Taking the continuous limit, the variance can be rewritten as
\begin{align}
{\rm var} \left(\hat{v}_{d_L-\bn'}\right)&=\frac{3}{\left(d^{(0)}_L\right)^2 N_{\rm tot}}\label{eq:var_final}\\
&\times \int\!\! {\rm d}r\, p(r,\rho_*)\left[\left\langle \Delta d_L^2(r) \right\rangle+\bar{d}^{\,2}_L(r)-\left(d^{(0)}_L\right)^2 \right]\, ,\nonumber
\end{align}
where $p(r, \rho_*)$ is the radial distribution of detected sources integrated over masses. 
In Fig.~\ref{fig:vardL_plots}, we plot the variance~\eqref{eq:var_final} for BBHs as a function of $N_{\rm tot}$ for different relative errors on the luminosity distance measurements, ranging from 10\% to 100\% (and constant in $r$). We use the radial distribution of sources obtained from our simulated catalogues (see Sec.~\ref{sec:sim} for more detail). We see that both the shot noise contribution and the distance-uncertainty contribution scale as $1/N_{\rm tot}$. The relative importance of the two terms depends therefore only on the uncertainty on $d_L$. For an uncertainty of up to 20\%, the shot noise contribution completely dominates in the variance. If the uncertainty reaches 50\% however, its contribution to the variance is not negligible anymore and the SNR is degraded. Similar results are obtained for BNSs, see Fig.~\ref{fig:vardL_plots_BNS} in Appendix~\ref{app:variance}. The only difference comes from the radial distribution of sources, which differs for BBHs and BNSs and leads to a slightly larger variance for BBHs (due to a wider redshift range of observation for BBHs, that can be observed at higher redshift than BNSs). In the following we will assume a 20\% uncertainty on the measurement of $d_L$ as a proxy of typical distance uncertainties for GW events \citep{2020ApJ...896L..32C,2022arXiv220702771I}, meaning that we are in the regime where shot noise completely dominates.

\subsection{Chirp mass}
\label{sec:estimator_mass}

The same procedure can be applied to the chirp mass. We first build an observable equivalent to~\eqref{dlObservable},
\begin{equation}
    v_{{\cal M}-\bn'}\equiv -3\int \frac{ {\rm d}\Omega}{4 \pi}\ \left(\bn\cdot\bn'\right) {\cal D M}(\bn)\simeq \alpha_{\cal M}  \frac{v_0}c \cos{\theta'}\,.
    \label{MObservable}
\end{equation}
We repeat the process of dividing the detected observations in sky angular pixels, as in Eq.~\eqref{eq:dLstat}, to associate the following statistical estimator to $ v_{{\cal M}-\bn'}$:
\begin{equation}\label{eq:Mstat}
    \hat{v}_{\mathcal{M}-\bn'}\equiv \frac{-3}{\hat{\mathcal{M}} N_{\rm sky} }\sum_{i=1}^{N_{\rm sky}}\sum_{ j=1}^{N_{\rm det}^i} \frac{(\mathcal{M})^i_j}{N_{\rm det}^i} (\bn_i\cdot\bn')\,,
\end{equation}
with
\begin{align}
\hat{\mathcal{M}}\equiv \frac{1}{N_{\rm sky}}\sum_{i=1}^{N_{\rm sky}}\sum_{j=1}^{N_{\rm det}^i}\frac{(\mathcal{M})^i_j}{N_{\rm det}^i}\,.
\end{align}

As for the luminosity distance, this estimator is found to be unbiased assuming that shot noise and uncertainty in the measurement of the chirp mass are subdominant compared to the mean. 
For the variance, we generalise the computation of Sec.~\ref{sec:estimator_distance} and Appendix~\ref{app:variance} done for the luminosity distance to have data not only distributed in $N_r$ radial bins, but also in $N_m$ bins of the two source masses, $m_{1, 2}$, which constitute the binary. This is needed since, unlike the luminosity distance, the redshifted chirp mass function depends on all $(r, m_{1,2})$. As for the luminosity distance, we neglect subdominant contributions from the observer velocity in the variance. We obtain 
\begin{align}
{\rm var}\left(\hat{v}_{{\cal M}-\bn'}\right)&=\frac{3 N_{\rm sky}}{\big({\mathcal{M}^{(0)}}\big)^2N^2_{\rm tot}}\Bigg\{\sum_{k=1}^{N_r}\sum_{\ell=1}^{N_m}  \langle(\Delta {\cal M}_{k \ell})^2 \rangle  \bar{\cal{N}}_p^{k\ell}   \label{varMSums}\\ 
+&\sum_{k=1}^{N_r}\sum_{\ell=1}^{N_m}  (\bar{\cal M}_{k \ell})^2 \bar{\cal{N}}_{p}^{k\ell}-  \frac{1}{\bar{N}_{\rm det}}\Bigg[\sum_{k=1}^{N_r}\sum_{\ell=1}^{N_m}\bar{\cal M}_{k \ell}\bar{\cal{N}}_p^{k\ell}\Bigg]^2\Bigg\}\,,\nonumber
\end{align}
where the $k$-index is the radial bin index, while the $\ell$-index is the index over the various $\mm$ bins. $\bar{\cal M}_{k \ell}$ is the expected redshifted chirp mass in such a bin, while $\bar{\cal{N}}_p^{k\ell}$ is the expected number of detected events in that bin, i.e.\ such that $\bar{N}_p^k=\sum_{\ell=1}^{N_m}\bar{\cal{N}}_p^{k\ell}$. The quantity $\Delta {\cal M}_{k \ell}$ denotes the typical uncertainty in the measurement of the chirp mass in a radial bin centered at $r_k$ and in the mass bin labelled by $\ell$.

\begin{figure}
    \centering
    \includegraphics[width=0.4\textwidth]{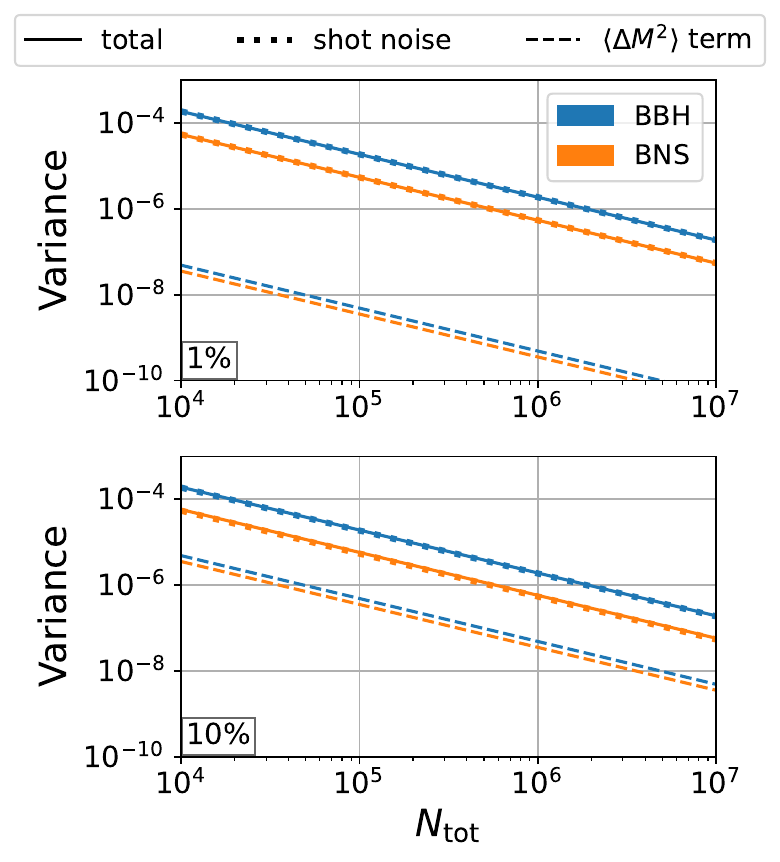}
    \caption{Different contributions to the variance of the chirp mass estimator, for a relative measurement error $\langle \Delta M^2(r, m_{1, 2})\rangle / \bar{M}^{2}(r, m_{1, 2}) = 1\% \text{ (top)}, 10\% \text{ (bottom)}$. In any case, shot noise is the fully dominant variance contribution, the measurement error giving a smaller contribution by orders of magnitude.}
    \label{fig:varM_plots}
\end{figure}

Taking the continuous limit, the variance \eqref{varMSums} becomes
\begin{align}
{\rm var}\left(\hat{v}_{{\cal M}-\bn'}\right)&=\frac{3}{\big({\mathcal{M}^{(0)}}\big)^2
N_{\rm tot}}\int\mathrm dr\,\mathrm d\mm\ p(r, \mm, \rho_*)  \label{varM} \\ 
&\times \left[\langle(\Delta\mathcal M)^2(r, \mm)\rangle+\bar{\mathcal M}^2(r, \mm)-\left({\mathcal{M}^{(0)}}\right)^2
\right]\,. \nonumber   
\end{align}
As for the luminosity distance, we see that the variance contains two contributions: one from measurement uncertainty of the chirp mass, and the second one from shot noise. In addition to the fact that shot noise affects the radial distribution of events in a given solid angle, it will also affect the distribution in masses $m_{1,2}$. This effect is encoded in the dependence of the distribution function $p$ on the masses and it will also generate fluctuations in the mean chirp mass that can mimic a dipole.  In Fig.~\ref{fig:varM_plots} we plot the different contributions as a function of $N_{\rm tot}$ assuming a 10\% uncertainty in the measurement of the chirp mass (expected for BBHs) and a 1\% uncertainty (expected for BNSs), see~\cite{2022arXiv220702771I}. We see that in both cases shot noise completely dominates over mass uncertainty. Comparing the shot noise contribution for BBHs and BNSs, we see that it is significantly smaller for BNSs, by a factor 3. This is due to the fact that BNSs have a narrower mass range than BBHs: the mass distribution of BNSs only span 2$M_{\sun}$, while BBHs are observed between 5$M_{\sun}$ and 100$M_{\sun}$. Since shot noise changes the mass distribution of events, it has more impact in the second case. For example, a shot noise fluctuation generating one more event at the higher end of the mass range will more drastically affect the mean mass of BBHs than the mean mass of BNSs. 

\subsection{Number counts}
\label{sec:estimator_count}

The dipole estimator for the number counts and its variance have been derived in detail in~\citet{Mastrogiovanni:2022nya}. Here we just state the final result for completeness:
\begin{align}
    v_{N-\bn'} &\equiv  \frac{3}{2} 
    \int \frac{\dd \Omega}{4\pi} \left(\bn\cdot\bn'\right) {\cal D}N(\bn) \simeq \alpha_{N} \frac{v_0}{c} \cos\theta'\, ,
    \label{CountObservable}\\
    \hat v_{N-\bn'} &= \frac{3}{2N_{\rm tot}} \sum_{i = 1}^{N_{\rm sky}} N_{\rm det}^i  (\bn_i\cdot\bn')\, .
\end{align}
The variance of the number count estimator is only due to shot noise and is given by
\begin{equation}
    {\rm var}\left(\hat v_{N-\bn'}\right) = \frac{3}{4 N_{\rm tot}}\, .
    \label{MeanVarCount}
\end{equation}

\subsection{Covariance of the estimators}

In order to use the distance, mass and number counts estimators together to measure the dipole from the same population of GW sources, it is necessary to compute their covariance. Below we show that the distance and mass estimators are correlated, but that these estimators are both uncorrelated with the number count estimator. Naturally, the estimators for two different populations of sources (for example BBHs and BNSs) are uncorrelated.

\subsubsection{Mass-distance covariance}
The covariance is calculated following the same steps as for the variance in Secs.~\ref{sec:estimator_distance} and~\ref{sec:estimator_mass}. Here we assume that the uncertainty in the measurement of the chirp mass is uncorrelated with the uncertainty in the measurement of the distance. The chirp mass is mainly measured from the phase, whereas the luminosity distance is measured through the amplitude of the wave. Hence, this is certainly a good approximation for BNSs, which have a long inspiral in the detectors' band and thus a good phase determination. Moreover, inspection of typical BBHs signals shows that the $d_L-{\cal M}_c$ entries of the normalized Fisher matrix are generally smaller than the diagonal ones, which we have already shown to give subdominant contributions to the total noise (see Figs.~\ref{fig:vardL_plots} and~\ref{fig:varM_plots}). This justifies the assumption of uncorrelated chirp mass and distance measurements uncertainties as well for the BBH type of sources. 

We obtain for the covariance
\begin{align}
{\rm cov}\left(\hat{v}_{d_L-\bn'},\hat{v}_{{\cal M}-\bn'} \right)&=\frac{3 N_{\rm sky}}{\mathcal{M}^{(0)}_{\phantom{L}}d^{(0)}_L N_{\rm tot}^2}\left\{\sum_{k=1}^{N_r}\sum_{\ell=1}^{N_m} \bar{\cal M}_{k \ell} (\bar{d}_L)_k  \bar{\cal{N}}_p^{k\ell}\right.\nonumber\\
&\left.-\frac{1}{\bar{N}_{\rm det}}\left[\sum_{k=1}^{N_r}\sum_{\ell=1}^{N_m}\bar{\cal M}_{k \ell}\bar{\cal{N}}_p^{k\ell}\right]\left[\sum_{j=1}^{N_r}(\bar{d}_L)_j\bar{N}_p^j\right]\right\}\,.
\end{align}
Taking the continuous limit we obtain
\begin{align}
  {\rm cov}\left(\hat{v}_{d_L-\bn'},\hat{v}_{{\cal M}-\bn'} \right)&=\frac{3}{ \mathcal{M}^{(0)}_{\phantom{L}} d^{(0)}_L N_{\rm tot}}
  \int\mathrm dr\,\mathrm d\mm\,p(r,\mm,\rho_*)\nonumber\\
  &\times \Big[\bar{\mathcal M}(r, \mm)\bar d_L(r)- \mathcal{M}^{(0)}_{\phantom{L}}d^{(0)}_L\Big]\,\label{CovMdContinuous}.
\end{align}
From Eq.~\eqref{CovMdContinuous} we see that the two estimators are correlated through shot noise. Indeed, if in a given solid angle there are more events closer to the observer, the mean distance in that solid angle will be smaller than on average, and the mean chirp mass as well. Hence a shot noise fluctuation that can mimic a dipole in one estimator will automatically mimic a dipole in the other estimator as well.

\subsubsection{Count-distance and count-mass covariance}
\label{sec:covariance}

We start by computing the covariance between the number counts and the distance estimators. Since shot noise is uncorrelated with the measurement uncertainty in the distance, we obtain, following the same steps as previously
\begin{align}
    & {\rm cov}\left(\hat{v}_{d_L-\bn'},\hat{v}_{N-\bn'} \right)=\frac{9}{2}\frac{1}{d^{(0)}_L N_{\text{tot}}N_{\text{sky}}}\sum_{i=1}^{N_{\text{sky}}}\sum_{a=1}^{N_{\text{sky}}}\sum_{\ell=1}^{N_r}({\bf{n}}_i\!\cdot\! {\bf{n'}})({\bf{n}}_a\!\cdot\! {\bf{n'}}) \nonumber\\
    &(\bar{d}_L)_{\ell} \bar{N}_p^{\ell}\left[\sum_{q=1}^{N_r}\left\langle \frac{\Delta N_p^{aq}}{\bar{N}_{\text{det}}}\frac{\Delta N_p^{i\ell}}{\bar{N}_p^{\ell}} \right\rangle -\sum_{q=1}^{N_r}\sum_{t=1}^{N_r}\left\langle   \frac{\Delta N_p^{aq }}{\bar{N}_{\text{det}}}\frac{\Delta N_p^{i t}}{\bar{N}_{\text{det}}}\right\rangle \right]\,.\label{eq:covdLN}
\end{align}
Using that the number of detected events follows a Poisson distribution, we can easily show that the two terms in Eq.~\eqref{eq:covdLN} exactly vanish. A similar calculation shows that the covariance between mass and number counts also vanish. This is not surprising: contrary to the distance and mass estimators, the number count estimator is not sensitive to the radial distribution of events. Only fluctuations in the total number of events in a solid angle can mimic a dipole in the number count. On the contrary, the mass and distance estimators are not sensitive to the total number of events in a solid angle (since we divide by the total number of events to obtain the mean) but rather to their radial distribution. As a consequence, a shot noise fluctuation that mimics a dipole in the number count estimator does not necessarily mimic a dipole in the mass and distance estimator, and vice versa.

\subsection{Optimal combination}
\label{sec:optimal}

In the case where there are no threshold effects, the six estimators (three for BBHs and three for BNSs) are estimators of the same quantity: $v_0/c\cos\theta'$. We can therefore look for an optimal estimator of this quantity, i.e.\ an estimator that would maximise the signal-to-noise ratio. In the following we drop the dependence in $\bn'$, since the optimal estimator can be defined in exactly the same way for each value of $\bn'$. 

As a first step, we build combinations of estimators that are independent. Since only the distance and the mass estimators are correlated for the same population of sources, we simply need to diagonalise that part of the covariance matrix. We obtain two new estimators (per population) given by
\begin{equation}
\label{eq:optimalest}
    \hat{v}_{\pm}=\frac{X^\pm\hat{v}_{d_L}+\hat{v}_{\cal M}}{1+X^\pm}\,,
\end{equation}
with
\begin{align}
X^{\pm} = &\frac{1}{2{\rm cov}\left(\hat{v}_{d_L},\hat{v}_{\cal M} \right)}\times\Bigg\{{\rm var}\left(\hat{v}_{d_L}\right)-{\rm var}\left(\hat{v}_{\cal M}\right) \nn\\
&\pm\sqrt{\rbr{{\rm var}\left(\hat{v}_{\cal M}\right)-{\rm var}\left(\hat{v}_{d_L}\right)}^2+4\, {\rm cov}^2\left(\hat{v}_{d_L},\hat{v}_{\cal M} \right)}\,\Bigg\}\,.
\end{align}
The denominator in Eq.~\eqref{eq:optimalest} insures that the new estimators have mean $v_0/c\cos\theta'$.

We can now build an optimal estimator of $v_0/c\cos\theta'$ by linearly combining the independent estimators in the following way
\begin{equation}
\hat{v}_{\rm ideal}\equiv\frac{\sum_{i=1}^6 c_i\hat{v}_i}{\sum_{i=1}^6 c_i}\,,\quad c_i=({\rm var}\left(\hat{\nu}_{i}\right))^{-1}\,,
\label{eq:ideal}
\end{equation}
with $i$ labeling all the possible observables (number count, plus and minus combinations) for both BBHs and BNSs. This combination is the one that maximizes the SNR (in every direction ${\bf n}'$), defined as
\begin{align}
    ({\rm SNR})^2=\frac{\langle\hat{v}_{\rm ideal}\rangle^2}{{\rm var}\left(\hat{v}_{\rm ideal}\right)}\,.
    \label{SNR}
\end{align}
Note that the normalisation in Eq.~\eqref{eq:ideal} is there to keep the condition $\langle\hat{v}_{\rm ideal}\rangle=v_0/c \cos\theta'$, but it is otherwise irrelevant.

If threshold effects are relevant, the mass, distance and number count estimators are not estimators of the same quantity, since the respective $\alpha$'s are different. In this case, we need to divide each estimator by its respective $\alpha$ before combining them. Since the $\alpha$'s can be determined only with limited precision (using population models, see Sec.~\ref{sec:alpha} for more detail), this would induce additional contributions to the variance of the optimal estimator and degrades the SNR. In Sec.~\ref{sec:Fisher_erroralpha}, we quantify this effect using a Fisher analysis.

\section{Measurement of the dipole from synthetic catalogues of GW sources}
\label{sec:sim}

To test our estimators we build synthetic catalogues of BBHs and BNSs and use our estimators on these simulated events. We compare the measurement with our theoretical modelling for the signal and the variance.

\subsection{Simulating BBHs and BNSs}
\label{sec:catalogues}

To build catalogues of BBH and BNS sources, we draw source frame masses for BBHs and BNSs from the same probabilistic models used in~\citet{Mastrogiovanni:2022nya} (see Appendix~A therein). These mass models are consistent with current BBH and BNS detections \citep{2022arXiv220702771I}. The redshift distribution of GW sources is determined by the merger rate model. In our simulation, the merger redshift is distributed according to 
\begin{equation}
    p(z)\propto \left [1+(1+z_p)^{-\gamma-k}\right] \frac{(1+z)^\gamma}{1+\left(\frac{1+z}{1+z_p}\right)^{\gamma+k}} \frac{1}{1+z} \frac{\de V_c}{\de z},
\end{equation}
where $\gamma, k$ and $z_p$ are parameters that control the merger rate model, while $\frac{\de V_c}{\de z}$ is the differential of the comoving volume. As fiducial values, we take $\gamma= 2.7$, $k=3$ and $z_p=2$.  The distribution of $\cos\iota$ is chosen to be uniform. Once a set of BBHs and BNSs is drawn, we add the effect of the observer velocity. Aberration is included by shifting the angular position by $\theta'=\theta-v_0/c\sin\theta$, where $\theta$ is the angle between the source position and the observer velocity. The luminosity distance and chirp mass are modified according to Eqs.~\eqref{dLv0} and~\eqref{Mv0} respectively. We produce two copies of both the BBHs and BNSs catalogue, one with the ``CMB value'' of the observer velocity, $v_0/c=1.2\cdot 10^{-3}$ and the other one with the ``AGN value'' of the observer velocity, which is 5 times larger: $v_0/c=6\cdot 10^{-3}$.

We can then calculate the SNR of each event using Eq.~\eqref{eq:SNR1}. We consider a network of XG detectors including ET \citep{2010CQGra..27s4002P} and two CE \citep{PhysRevD.91.082001,Reitze:2019iox}. The power spectral density of ET is set to the one used in \citet{2022arXiv220702771I,Mastrogiovanni:2022nya}, while for CE we take the power spectral density taken from the CE consortium \footnote{\url{https://cosmicexplorer.org/sensitivity.html}}. If the SNR from this network exceeds a detection threshold of $\rho_*=9$, we label the binary as detected.

Finally, we add an extra step in the simulation to mimic the fact that we will not be perfectly able to measure the sky location, luminosity distance and redshifted chirp mass of the source. Once a binary is detected, we do not save its true values for the sky position, distance and chirp mass but instead, we register a scattered value around the true one. We include Gaussian scatter of the sky location by $3$ degrees around the true sky position and of the luminosity distance by 20$\%$ of its true value. For the chirp mass, we use a scattering of $1\%$ for BNSs and of $10\%$ for BBHs. These are typical errors that we might expect to obtain with XG detectors \citep{2022arXiv220702771I}. {Note that the 20\% uncertainty on the luminosity distance implies that shot noise is the dominant contribution, as can be seen from Figs.~\ref{fig:vardL_plots} and~\ref{fig:vardL_plots_BNS}.}

\begin{figure}
    \centering
    \includegraphics[scale=0.55]{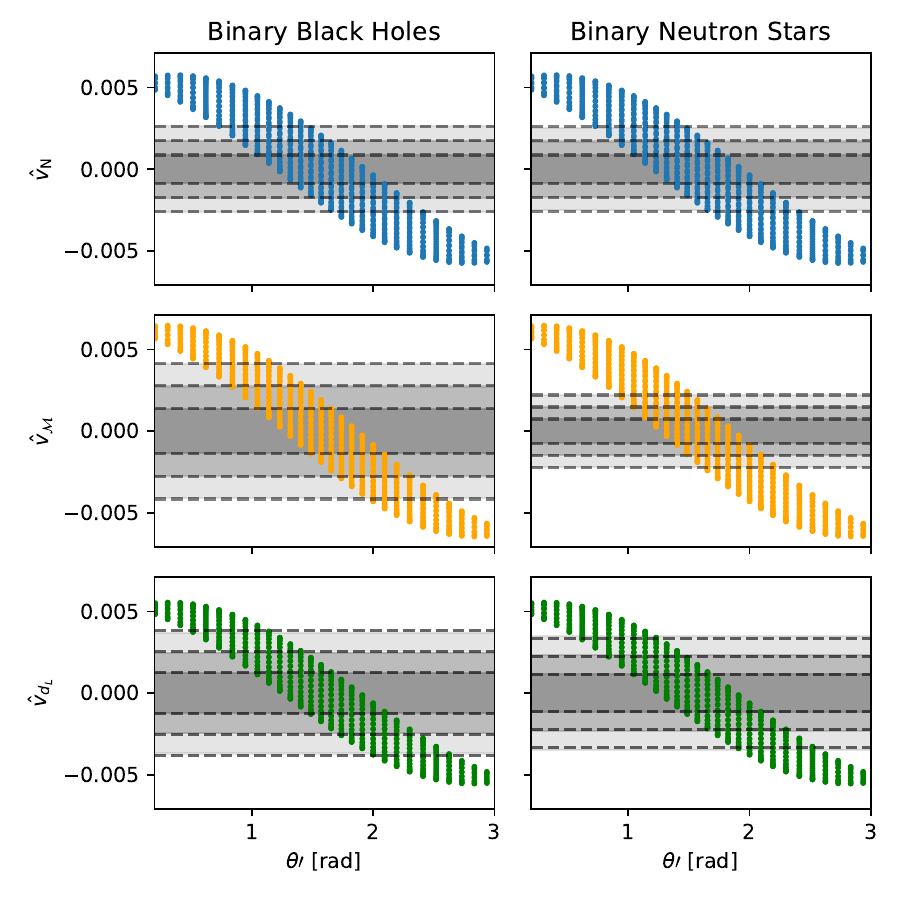}
    \caption{Value of the dipole estimators as a function of the angle between the true dipole direction and a chosen direction $\bn'$. The plots are generated using the AGN value of the observer velocity, and with $10^6$ BBH detections (left panel) and $10^6$ BNS detections (right panel). The grey shaded areas are the $1,2,3 \sigma$ values associated with the shot noise and measurement uncertainties in the absence of dipole, obtained by shuffling the sources isotropically over the sky one hundred times. The horizontal dashed lines mark the theoretical expectations for the variance obtained in Sec.~\ref{sec:estimators}.    
    \textit{Top plot:} Number count estimator. \textit{Middle plot:} Chirp mass estimator. \textit{Bottom plot:} Luminosity distance estimator.} 
    \label{fig:projections}
\end{figure}

The luminosity distances, redshifted chirp masses and sky positions are then used to compute the estimators using Eqs.~\eqref{eq:dLstat},~\eqref{eq:Mstat} and~\eqref{CountObservable}. The results are shown in Fig.~\ref{fig:projections} for the case of the AGN observer velocity. We show the values of the estimator obtained for $10^6$ BBH detections (left panel) and $10^6$ BBN detections (right panel) as a function of the angle $\theta'$ between the true direction of the dipole and a chosen direction $\bn'$. The spread in the signal comes from the fact that for a given angle $\theta'$ we have pixels in different azimuthal directions, which give slightly different values for the dipole estimator (due to shot noise and measurement uncertainties). At the equator there are clearly more azimuthal pixels than at the pole (exactly at the pole there is just one), leading to a larger spread at the equator.

As discussed in~\cite{Mastrogiovanni:2022nya}, for BBHs threshold effects do not contribute to the amplitude of the dipole, since all events are above the threshold (see Fig.~2 of~\cite{Mastrogiovanni:2022nya}). As a consequence, at the maximum, i.e.\ when $\bn'$ coincides with the direction of the observer velocity, the estimators are roughly equal to $v_0/c=6\cdot 10^{-3}$. The peak of the dipole estimators could be slightly shifted from the true position due to the variance. For BNSs, we see that the amplitude of the dipole at the peak is also very close to the observer velocity. This is due to the fact that, as we will show in Sec.~\ref{sec:alpha}, the amplitude of threshold effects are actually small for BNSs, below 10\%, which is smaller than the spread in the signal.

Fig.~\ref{fig:projections} also reports the $1,2,3\sigma$ fluctuations (grey areas) of the estimators due to shot noise and measurement uncertainties on the luminosity distance and chirp mass. The fluctuation levels are generated by shuffling GW detections isotropically over the sky  a hundred times. We see that the chirp mass from BNSs is the estimator with smaller variance, consistent with the theoretical results of Sec.~\ref{sec:estimators}.  For all estimators, the fluctuation levels obtained through sky shuffling agree with the theoretical calculation of the variance from Eqs.~\eqref{eq:var_final},~\eqref{varM} and~\eqref{MeanVarCount}, that are indicated with dashed lines on the plot (see Sec.~\ref{sec:comp_cov} for a more detailed comparison). When the estimator values exceed a certain noise threshold, the cosmic dipole can be detected (see Sec.~\ref{sec:detectability} for a more in-depth discussion on detectability).

\subsection{Comparison of the theoretical variance and covariance with simulations}
\label{sec:comp_cov}

\begin{figure*}
    \centering
    \includegraphics[scale=0.6]{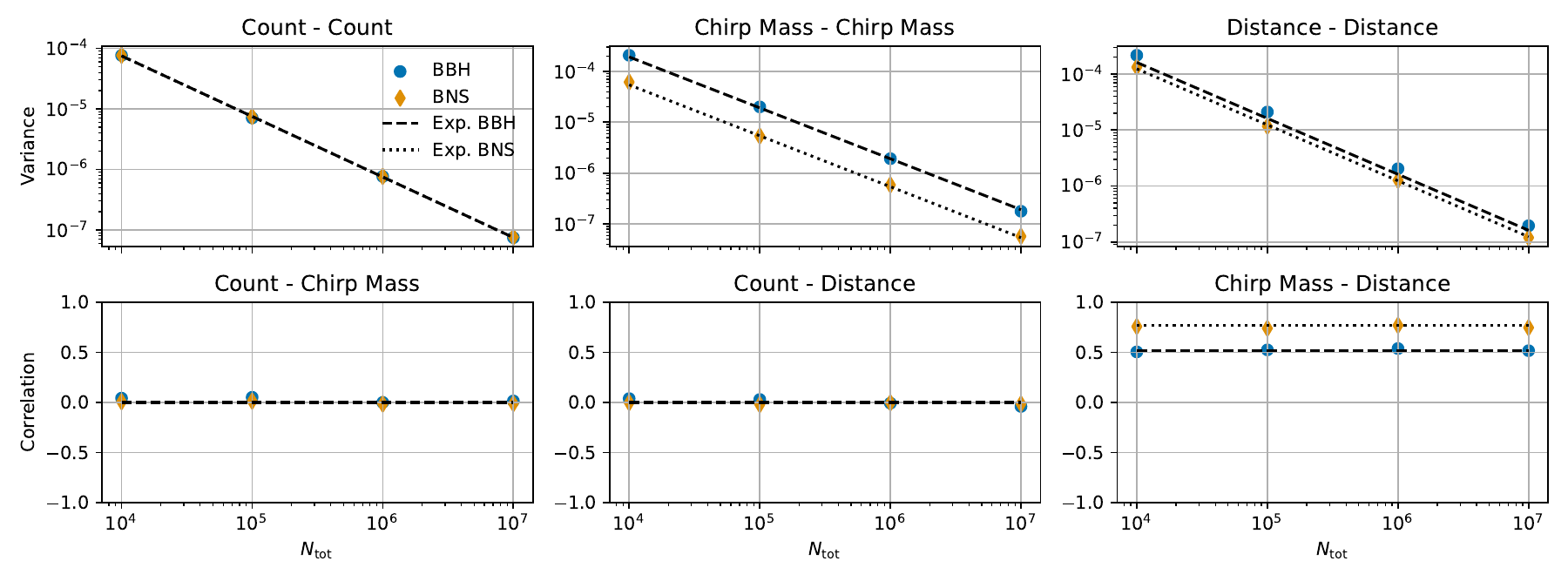}
    \caption{The plots report the variance (first row) and correlation (second row) of the number count, mass, and distance dipole estimators as a function of the number of detections $N_{\rm tot}$. The variances and correlations are obtained by reshuffling 2'000 times isotropically a population of GW sources. The blue circles indicate the values obtained for the BBH population while the orange diamonds the values for the BNS population. The black lines (dashed for BBHs and dotted for BNSs) indicate the theoretical variance calculated in Sec.~\ref{sec:estimators}. We find that the maximum deviation from the expected value of zero for the \textit{number count - chirp mass} and \textit{number count - luminosity distance} is 0.05,  arising from the limited number of sky shufflings.} 
    \label{fig:covariance_matrix}
\end{figure*}

The first sanity check that we perform is to see if our predictions of the variance and covariance agree with the numerical simulations. As explained above, the numerical variance and covariance are simply obtained by shuffling the sources isotropically over the sky. This removes the true dipole signal, meaning that the remaining fluctuations are due to shot noise and measurement uncertainties of the luminosity distance and chirp mass. Here we use 2'000 sky shufflings to obtain an accurate numerical estimate of the variance and covariance.

Theoretically, the variance of the number count estimator is given by Eq.~\eqref{MeanVarCount}, and simply depends on the total number of detected events.  On the other hand, the variance of the mass and distance estimators as well as the covariance between them has to be computed from the integrals \eqref{eq:var_final}, \eqref{varM} and \eqref{CovMdContinuous}, which requires a model for the $(r, \mm)$ source distributions. To estimate this, we numerically generate GW events following reference mass and redshift distributions models, and approximate the integrals in~\eqref{eq:var_final}, \eqref{varM} and \eqref{CovMdContinuous} by sampling those distributions. The sampled number of detections is increased until we reach numerical convergence. 

The top plots of Fig.~\ref{fig:covariance_matrix} show the simulated variance compared with the theoretical one, for BBHs and BNSs, as a function of the total number of events $N_{\rm tot}$. The agreement between the simulated and theoretical variances is excellent. As expected from Eqs.~\eqref{eq:var_final},~\eqref{varM} and~\eqref{MeanVarCount} the variances scale as $N_{\rm tot}^{-1}$. For the number count estimator, the variance is the same for BBHs and BNSs: it depends only on the total number of events. For the chirp mass estimator, on the other hand, the variance for the BNSs is smaller by a factor of 3. As already discussed in Sec.~\ref{sec:estimator_mass}, this is due to the narrower mass range of BNSs. For the luminosity distance estimator we see that the variance is also slightly smaller for BNSs. Again, this is due to the  slightly narrower radial distribution of BNS events compared to BBH events.

Comparing the different estimators, we see that the one with smaller variance is the mass estimator for BNSs. This is due to the fact that shot noise only affects the mass estimator by changing the radial distribution and $m_{1,2}$-distribution of sources. If the chirp mass would be constant in $m_{1,2}$ and in $r$, the last two terms in the second line of Eq.~\eqref{varM} would cancel each other. Although the chirp mass does depend on $m_{1,2}$ and $r$, the distribution of BNSs in masses and redshift is narrow enough for a reduction to occur. This leads to a shot noise contribution smaller than the one for the number count estimators, where there is no such reduction. For BBHs however, the wider range leads instead to an increased variance.

Finally, we see that the variance of the distance estimator is slightly larger than that of the number count both for BBHs and BNSs, due to the large variation of distance with $r$. The luminosity distance varies indeed as $(1+z)r$, while the redshifted chirp mass varies as $(1+z)$, i.e.\ significantly slower. 

Since the means of the three estimators are very similar (see Sec.~\ref{sec:alpha}), the mass estimator for BNSs is optimal in terms of SNR. On the other hand, BNSs are affected by threshold effects, meaning that the $\alpha$ coefficients need to be modelled if one wants to measure the observer velocity. Including the uncertainty in the modelling of the $\alpha$'s in the analysis generates an extra contribution to the variance, which needs to be accounted for. In Sec.~\ref{sec:Fisher} we quantify this effect and show that, despite it, the BNSs mass estimator strongly contributes to the constraints on the observer velocity.

The bottom panels of Fig.~\ref{fig:covariance_matrix} show the values of the correlations (i.e.\ the covariance divided by the square root of the respective variances) as a function of the number of BBH and BNS detections. As we can see from the plot, also in the simulations, we obtain that the number count estimator does not correlate with the mass and distance estimators. On the other hand, as expected, we find that the distance and mass estimators are positively correlated, in excellent agreement with the theoretical calculation. As explained before, this is due to the fact that the two observables are similarly sensitive to the radial distribution of sources.

\subsection{Detection efficiency of the dipole}
\label{sec:detectability}

\begin{figure*}
    \centering
    \includegraphics[scale=0.54]{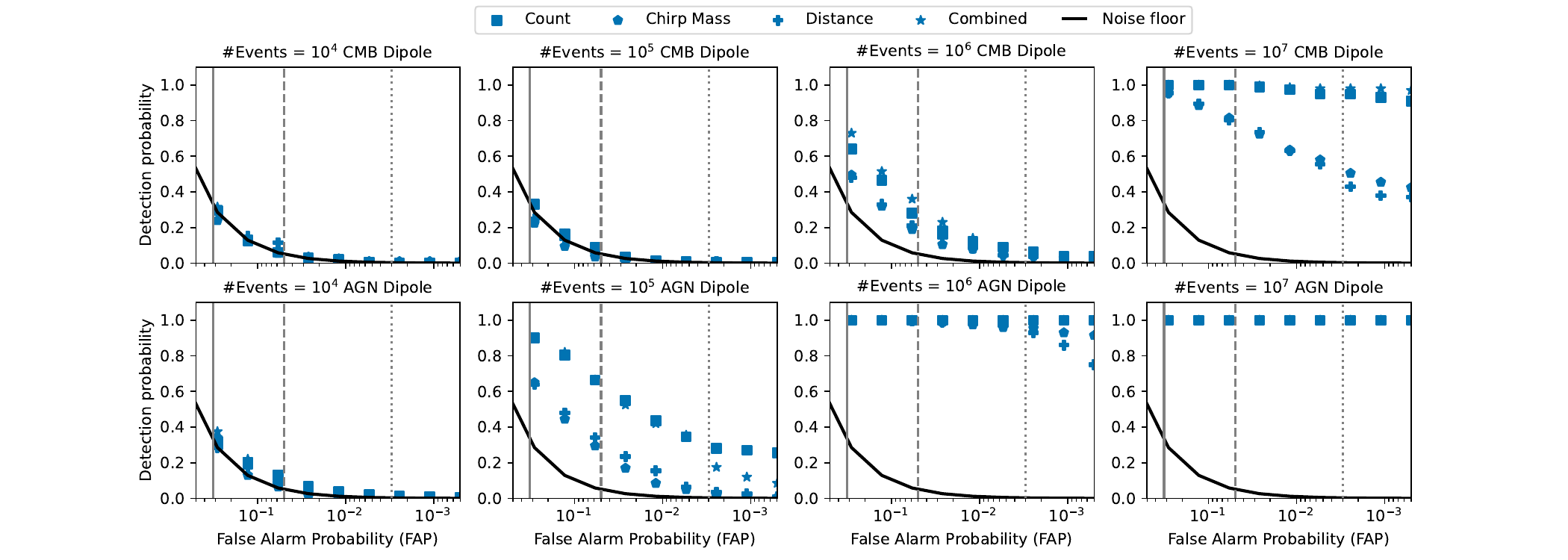}
    \caption{Detection probability for the cosmic dipole (vertical axis) versus false alarm probability (horizontal axis) for a population of BBH detections. The different columns consider different numbers of GW detections. The first row corresponds to a dipole consistent with the CMB cosmic dipole and the second to a dipole consistent with the AGN dipole. The different marker types (see legend) indicate the 3 estimators and their combination. The solid curve marks the detection probability/FAP relation in the case of fluctuations arising from an isotropic background. The detection probability is calculated by generating 200 population realisations while the FAP threshold is calculated using 2'000 noise realisations with the sky shuffling method. The vertical solid, dashed and dotted lines mark the $1,2,3 \sigma$ false alarm probabilities.}
    \label{fig:BBHs_detectionprob}
\end{figure*}

\begin{figure*}
    \centering
    \includegraphics[scale=0.54]{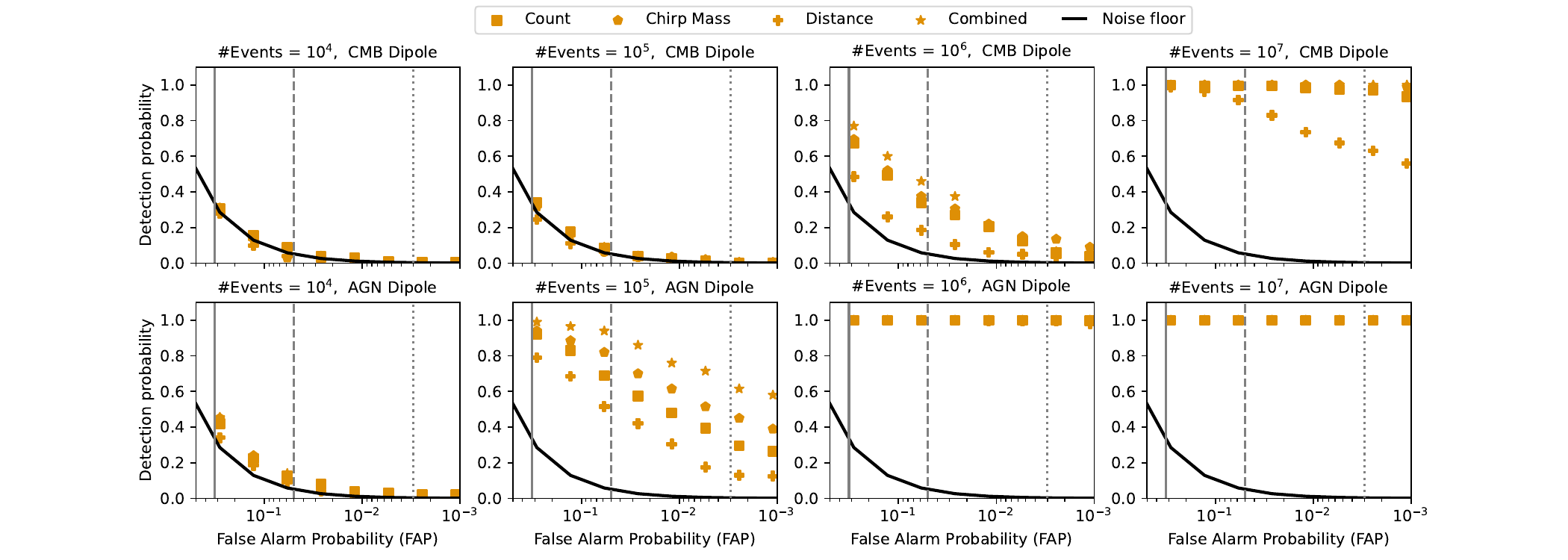}
    \caption{Same as Fig.~\ref{fig:BBHs_detectionprob}, but for BNSs instead of BBHs.
    }
    \label{fig:BNSs_detectionprob} 
\end{figure*}

\begin{figure*}
    \centering
    \includegraphics[scale=0.7]{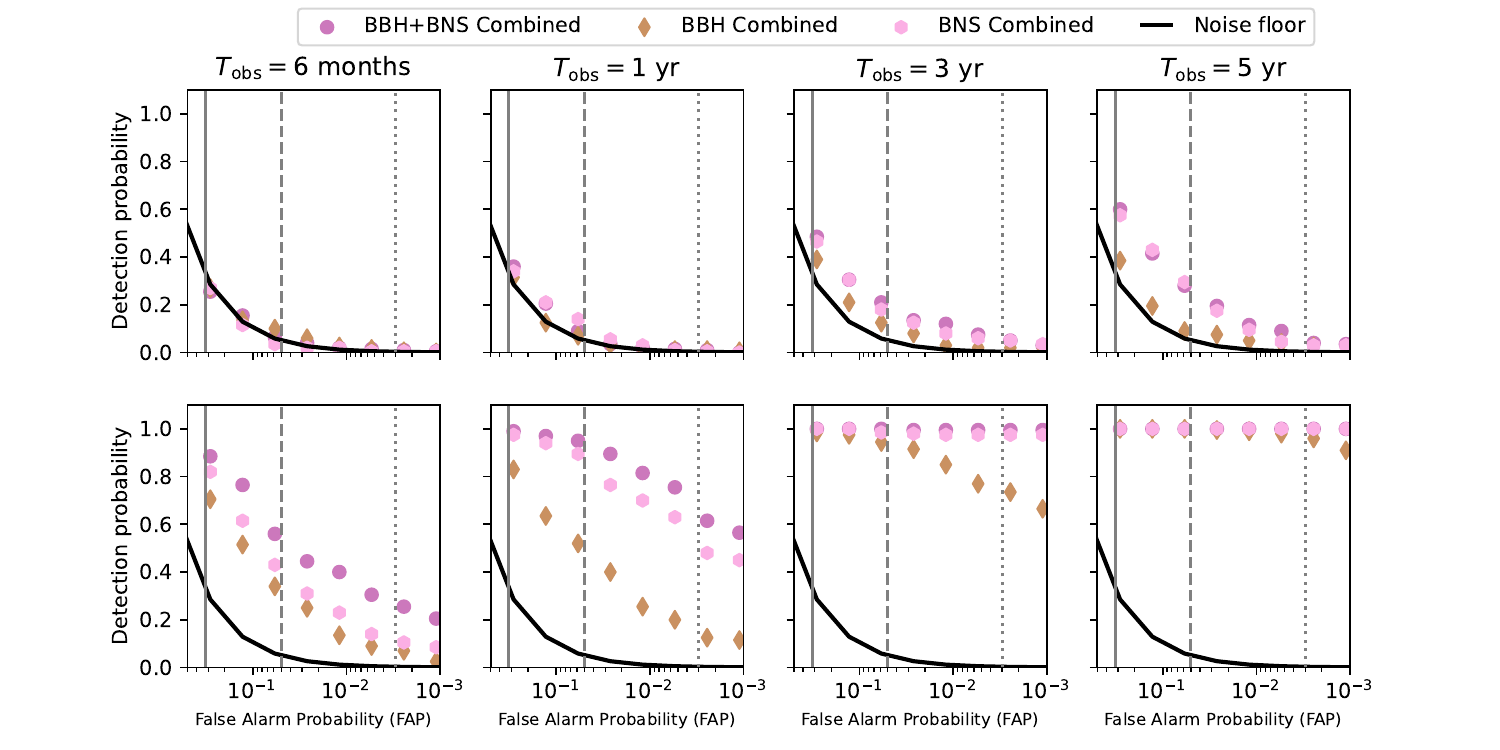}
    \caption{Detection probability for the cosmic dipole (vertical axis) versus false alarm probability (horizontal axis) using the combined estimator for a mixed population of GW sources. The different columns consider different observation times. Following \citet{2022arXiv220702771I}, for 1 year of observations we have taken $7.5\cdot 10^4$ BBH detections and $10^5$ BNS detections. The number of events expected in 6 months and in $3$ and $5$ years of observations are found by linearly scaling the previous fiducial values. The first row corresponds to a dipole consistent with the CMB cosmic dipole and the second to a dipole consistent with the AGN dipole. The solid curve marks the detection probability/FAP relation in the case of fluctuations arising from an isotropic background. The detection probability is calculated by generating 200 population realisations while the FAP threshold is calculated using 2'000 noise realisations obtained with the sky shuffling method. The vertical solid, dashed and dotted lines mark the $1,2,3 \sigma$ false alarm probabilities.
    }
    \label{fig:Tobs_detectionprob}
\end{figure*}

We now assess the detection efficiency of the dipole from the three estimators, for the BBH and BNS populations. For this we report in Figs.~\ref{fig:BBHs_detectionprob} and~\ref{fig:BNSs_detectionprob}, the detection probability versus false alarm probability (FAP) for several cases. The FAP identifies a threshold for the dipole detection and it is defined as the probability that a random fluctuation in the absence of a dipole (due to shot noise or measurement uncertainty), would result in a false positive. The detection probability is defined as the probability that, in the presence of a dipole, the estimator for the dipole detection would exceed the FAP threshold. For example, if we find a detection probability of 0.8 at an FAP of 0.32, it means that we have a probability of 80\% to detect the dipole at more than 1\,$\sigma$ significance. In the other 20\% of times, the dipole would be consistent with zero within the 1\,$\sigma$ error bars. However, noise alone gives one a 32\% percent chance of a detection.
 Clearly, the lower the FAP, the higher our detection confidence. On the other hand, a lower FAP requires the dipole signal to be larger in order to be detected, which therefore decreases the detection probability. In Figs.~\ref{fig:BBHs_detectionprob} and~\ref{fig:BNSs_detectionprob}, we also draw with a solid black curve the noise-dominated limiting case, where the FAP equals the detection probability, i.e.~where all the dipole detections are due to noise fluctuations alone.

As we can see from the plots, a dipole consistent with the CMB value  becomes detectable after collecting more than $10^6$ GW events, while a dipole consistent with the AGN value becomes detectable already after $10^5$ detections. This result is valid for both the BBH and BNS populations. 

Comparing the detection efficiencies of the individual estimators, we see that, in the BBH case (Fig.~\ref{fig:BBHs_detectionprob}), the number count estimator (square) is clearly the most efficient one. This directly follows from the amplitude of the variance, which as shown in Fig.~\ref{fig:covariance_matrix} is smaller for the number count estimator, due to the large mass and radial range of BBH events that increase the variance of the mass and distance estimators. In Fig.~\ref{fig:BBHs_detectionprob}, we also show the result obtained by combining the three estimators, accounting for the covariance between the mass and distance estimators. We see that in the case of BBHs, the combined estimator (stars) does not perform better than the number count estimator, meaning that the mass and distance estimators are irrelevant. Note that the fact that sometimes the combined estimator is below the number count estimator is due to numerical uncertainties, related to the fact that the variance and covariance are generated using only 2'000 sky shufflings.  

In Fig.~\ref{fig:BNSs_detectionprob} we show the results for BNSs. We clearly see that in this case the mass estimator has the highest detection efficiency, better than the number count estimator. This is directly related to the shot noise suppression discussed in Sec.~\ref{sec:comp_cov}. The distance estimator performs better for the BNSs than for the BBHs, due to the smaller redshift range spanned by BNSs, which also reduces the shot noise contribution with respect to the BBH case. Despite this, the efficiency of the distance estimator remains below that of the mass and number count. Combining the three estimators leads to a non-negligible gain in terms of detection efficiency~\footnote{Note that if we want to combine the estimators to extract the observer velocity $v_0/c$, we need to divide each estimator by its appropriate $\alpha$, as discussed in Sec.~\ref{sec:optimal}. On the other hand, if we just want to assess the detectability of the combined dipole, we do not need this extra step. In this case, the mean of the combined estimator does however not provide a measurement of $v_0/c$. }.

Finally, we also simulated a more realistic scenario for the detection of the cosmic dipole, where we do not perform separate analyses for BNSs and BBHs, but instead combine all the BBHs and BNSs detected in a given observing time $T_{\rm obs}$. Note that in this combination, we still apply the estimators separately on the BNS and BBH populations, since we want to preserve the fact that BNSs have a smaller mass range and radial range than BBHs. We consider that in one year, the network of ET + 2CE detectors would be able to  observe $7.5 \cdot  10^4$ BBHs and $10^5$ BNSs. In Fig.~\ref{fig:Tobs_detectionprob} we show the detection efficiency for the combined estimator for 6 months, 1 year, 3 years and 5 years of observations. On one hand, we find that, with 5 years of observations and a bit less than $10^6$ sources detected, we can detect the dipole with 1\,$\sigma$ significance, but it will be unlikely to reach a high significance of 3\,$\sigma$. On the other hand, we find that a dipole consistent with the AGN one could be detected with high significance already with one year of observations, thanks to the BNS population. Therefore, a non-detection of the cosmic dipole in the first year of XG detectors would automatically rule out the AGN value of the dipole, thus providing a strong indication for the presence of un-modelled systematic in the AGN measurements.

\subsection{Modelling of the $\alpha$'s}
\label{sec:alpha}

\begin{table}
\centering
\caption{Expected values for the BNS $\alpha$ parameters, for a fiducial astrophysical population model. }\label{tab:BNSalpha}
\begin{tabular}{cccc}
\toprule
 &$\alpha_{N}$& $\alpha_{d_L}$ & $\alpha_{\cal M}$ \\
\midrule
 BNS & 1.08 & 0.94 & 0.98  \\ 
\bottomrule
\end{tabular}
\end{table}

\begin{figure}
    \centering
    \includegraphics[scale=0.5]{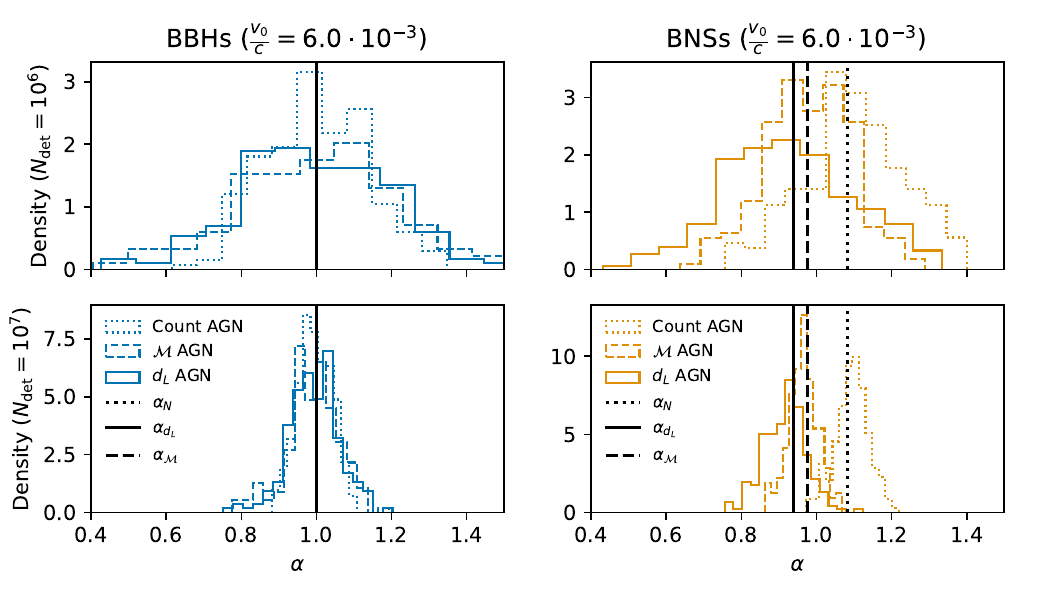}
    \caption{The plots indicate the distribution of the $\alpha$ parameters obtained for the 3 estimators with 200 population realisations. The vertical black lines indicate the theoretical prediction, see Table~\ref{tab:BNSalpha}. The histogram width is due to shot noise and distance and mass uncertainty, which vary over the 200 realisations. The first row of plots considers $10^6$ detections and the second row $10^7$ detections. The first column is for BBHs, while the second for BNSs. The plots are generated with the AGN value of the observer velocity.}
    \label{fig:alpha_AGN}
\end{figure}

\begin{figure}
    \centering
    \includegraphics[scale=0.5]{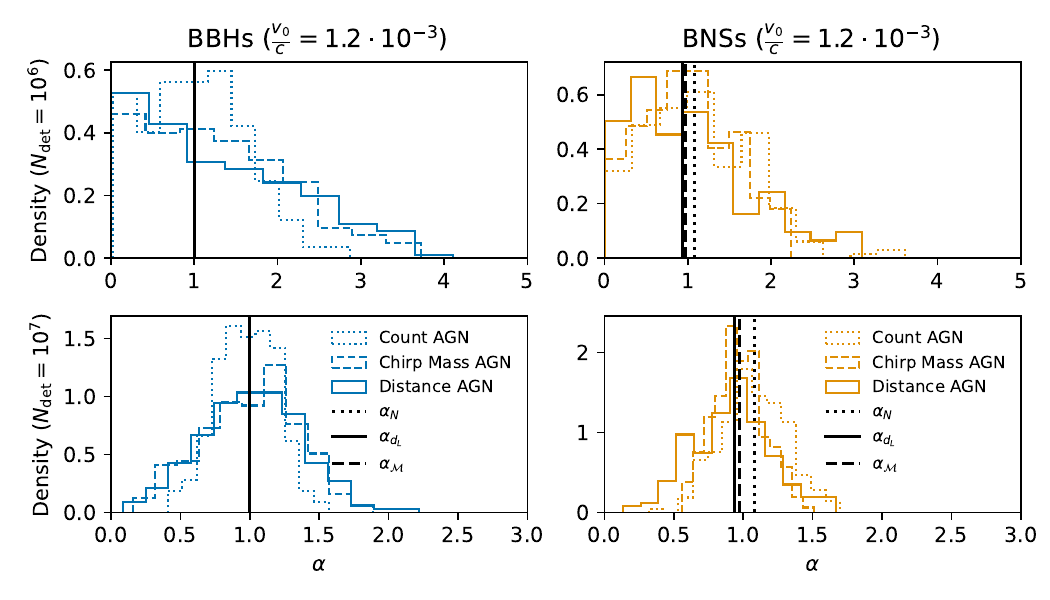}
    \caption{Same as Fig.~\ref{fig:alpha_AGN}, but with the CMB value of the observer velocity.}
    \label{fig:alpha_CMB}
\end{figure}

\begin{figure}
    \centering
    \includegraphics[scale=0.5]{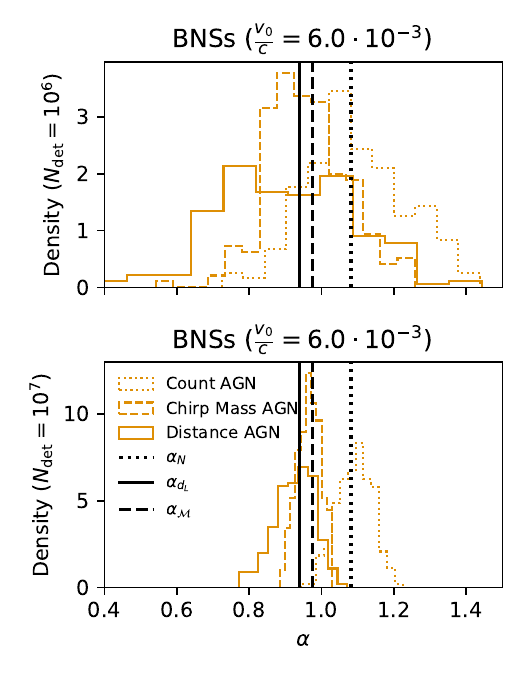}
    \caption{The plots indicate the distribution of the $\alpha$ parameters obtained for the 3 estimators with 200 population realisations. Each realisation also considers a random realisation of the merger rate model parameters $\gamma$ with an uncertainty of $10\%$ around its fiducial value of $\gamma=2.7$. The vertical black lines indicate the theoretical prediction from Table~\ref{tab:BNSalpha} for the $\gamma$ fiducial value. The histogram width is due to shot noise, distance and mass uncertainty, and also variation of the merger rate parameter. The top plot is for $10^6$ detections while the second one for $10^7$ detections. The plots are generated with the AGN value of the observer velocity.}
    \label{fig:alpha_ANG_BNS_pop}
\end{figure}

The results of the previous section show that the mass estimator for the BNSs is better than the count estimator, both for BNSs and BBHs, in terms of detectability of the dipole signal. However, BNSs (contrary to BBHs) are affected by threshold effects. In order to use the BNSs mass estimator to measure the observer velocity, it is necessary to have a modelling of the coefficients $\alpha$. 

The parameters $\alpha$ depend on the population of sources, through the parameters $s$ and $\mathcal{A}$ defined in Eqs.~\eqref{newdefhats} and~\eqref{newdefhatA}. We use the mass model defined in Appendix~A of~\cite{Mastrogiovanni:2022nya} to describe the population of BNSs. We then bin the events in $(r, \mm, \rho_*)$ around $\rho_* = 9$, and compute with finite differences the derivative with respect to $\rho_*$ to obtain $s$ through Eq.~\eqref{newdefhats}. This needs to be done for each $(r, \mm)$ bin, at $\rho_* = 9$. We then use interpolation in $r$ and $\mm$ to promote the binned $s$ values to a function.
The function $\cal A$ depends on $\cal{F}$, which quantifies the sensitivity of the detector. We first evaluate numerically $\cal F$ on a discrete set of points and then interpolate between them, allowing to attribute one $\calA$ value to each event. We can then estimate the integrals~\eqref{defalphad}, \eqref{defalphaM} and~\eqref{defalphaN} to obtain $\alpha_{d_L}, \alpha_{\cal M}$ and $\alpha_{N}$. 

The results are shown in Table~\ref{tab:BNSalpha}. We see that the values are very close to one in all three cases. Threshold effects seem to be slightly suppressed in the mass and distance case with respect to the number count, probably due to the fact that the integrals~\eqref{defalphad} and~\eqref{defalphaM} vanish if $s$ and $\mathcal{A}$ are constant in $r$. 

Since we know the observer velocity in our synthetic catalogues of BNS events, we can use the dipole to measure the parameters $\alpha_{d_L}, \alpha_{\cal M}$ and $\alpha_{N}$ and compare this with our modelling. In reality, we would not be able to do that and the only quantity that can be measured is the product of $v_0/c$ and the respective $\alpha$. In Figs.~\ref{fig:alpha_AGN} and~\ref{fig:alpha_CMB} we show the values obtained for the $\alpha$'s assuming an observer velocity consistent with the AGN dipole and CMB dipole, respectively. For BBHs, the histograms are centered around 1, as expected. For BNSs, the histograms are slightly displaced due to threshold effects. The peak is in excellent agreement with the theoretical predictions for the $\alpha$'s. This is important for two reasons: first it shows that threshold effects are indeed small and should not spoil too much our measurement of the observer velocity. In particular, one of the goals of using GWs to measure the dipole is to determine if GWs are consistent with the AGN dipole or with the CMB dipole. Having threshold effects of the order of 10\% means that this test can be done robustly also using BNSs. Indeed, we can conclude that if we find a dipole consistent with the AGN one, it would be unlikely that this was due to very large threshold effects increasing the $\alpha$'s by a factor 5. Second, the plots show excellent agreement between the theoretical modelling and the measured $\alpha$'s. Since in practice the $\alpha$'s can only be modelled (and not measured), it is important to know that this can be done in a robust way. 

In Fig.~\ref{fig:alpha_ANG_BNS_pop}, we check the dependence of the $\alpha$'s on the population model. We vary the value of one of the parameter in the population model by 10\% around its fiducial value and compute the histograms for the $\alpha$'s. Comparing the width in the $\alpha$'s with the one only due to shot noise and measurement uncertainty (Fig.~\ref{fig:alpha_AGN}), we see that varying the model does not generate an additional spread in the $\alpha$'s. This means that the uncertainty from the choice of model is smaller than the variance of the dipole.

\section{Determination of the amplitude of the observer velocity}
\label{sec:Fisher}

Let us now use the Fisher formalism to estimate how well we can measure the amplitude of the velocity, $v_0/c$, by combining the different estimators. We consider two cases, the first one where the only unknown is the observer velocity, i.e.\ we assume that the parameters $\alpha_N,\alpha_{d_L}$ and $\alpha_{\cal{M}}$ are perfectly known; and the second one where these parameters are treated as free parameters, that can be determined (through additional measurements or through a theoretical modelling) with some uncertainty. In both cases, we assume that the direction of the dipole has already been determined by maximising the estimators $\hat{\nu}_{\bn'}$ with respect to $\bn'$.

\subsection{Known $\alpha$'s}
\begin{table*}
\centering
\caption{Error, $\sigma_{v_0/c}$, obtained from: the individual estimators (first 6 columns), combining the three BBH estimators, the three BNS estimators, all estimators, or using the top  3 combination of mass and number count estimator for BNSs and number count estimator for BBHs. In the top three rows, we consider a dipole consistent with the CMB one, while in the bottom three rows we consider a dipole consistent with the AGN one.}\label{tab:error_known}
\begin{tabular}{cccccccccccc}
\toprule
& &$\hat{\nu}_N^{\rm BBH}$ & $\hat{\nu}_{d_L}^{\rm BBH}$ & $\hat{\nu}_{\cal{M}}^{\rm BBH}$ & $\hat{\nu}_N^{\rm BNS}$ & $\hat{\nu}_{d_L}^{\rm BNS}$ & $\hat{\nu}_{\cal{M}}^{\rm BNS}$
& BBH & BNS & All & Top 3\\
\midrule
\parbox[t]{3mm}{\multirow{3}{*}{\rotatebox[origin=c]{90}{CMB}}}&$N_{\rm tot}=10^5$ &223\% &327\% &356\% & 206\%& 305\%& 194\% & 178\%& 139\%& 110\%& 119\% \\ 
&$N_{\rm tot}=10^6$ & 70\%& 103\%&113\% &65\% &96\% &61\% &56\% &44\% &35\%&  38\%\\
&$N_{\rm tot}=10^7$ & 22\%& 33\%& 36\%&21\% &30\% &19\% & 18\%& 14\%& 11\% & 12\%\\
\midrule
\parbox[t]{3mm}{\multirow{3}{*}{\rotatebox[origin=c]{90}{AGN}}}&$N_{\rm tot}=10^5$ & 46\%&67\% &73\% &42\% &63\% & 40\%& 36\%& 29\%& 23\%& 24\% \\ 
&$N_{\rm tot}=10^6$ &14\%& 21\% &23\% &13\% &20\% &13\% & 12\%&9\% &7\% &  8\%\\
&$N_{\rm tot}=10^7$ & 5\%&7\% & 7\%&4\% & 6\%& 4\%& 4\%&3\% &2\%  & 2\%\\
\bottomrule
\end{tabular}
\end{table*}

We use the Fisher formalism to estimate the uncertainty on $v_0/c$ obtained from the measurement of the dipole amplitude from the number count, chirp mass and luminosity distance of the BBH and the BNS populations. The signal contains therefore six measurements:
\begin{align}
S\equiv
\left(
\hat{v}^{\rm BBH}_{N}, 
\hat{v}^{\rm BBH}_{d_L},
\hat{v}^{\rm BBH}_{\cal{M}},
\hat{v}^{\rm BNS}_{N}, 
\hat{v}^{\rm BNS}_{d_L},
\hat{v}^{\rm BNS}_{\cal{M}}
\right)\, .  
\end{align}
The Fisher for the (unique) parameter $v_0/c$ is then given by
\begingroup
\renewcommand*{\arraystretch}{1.5}
\begin{align}
\mathcal{F}_{v_0/c}&=\sum_{ij}\frac{\partial S_i}{\partial {(v_0/c)}} {\rm cov}(S_i,S_j)^{-1}   \frac{\partial S_j}{\partial {(v_0/c)}}\\
&=\sum_{\substack{X= {\rm BBH},\\  {\rm BNS}}}\begin{pmatrix}\alpha^{\rm X}_N\\ \alpha^{\rm X}_{d_L}\\ \alpha^{\rm X}_{\cal{M}}
\end{pmatrix}
{\rm cov}_X^{-1}
\begin{pmatrix}\alpha^{\rm X}_N\\ \alpha^{\rm X}_{d_L}\\ \alpha^{\rm X}_{\cal{M}}\end{pmatrix}\, ,
\end{align}
\endgroup
where
\begingroup
\renewcommand*{\arraystretch}{1.8}
\begin{align}
\label{eq:covX}
{\rm cov}_X=
\begin{pmatrix}
{\rm var}\left(\hat{v}^X_{N}\right) & 0 & 0\\ 
0 & {\rm var}\left(\hat{v}^X_{d_L}\right) &{\rm cov}\left(\hat{v}^X_{d_L}, \hat{v}^X_{\cal{M}}\right)\\
0&{\rm cov}\left(\hat{v}^X_{d_L}, \hat{v}^X_{\cal{M}}\right)&{\rm var}\left(\hat{v}^X_{\cal{M}}\right)
\end{pmatrix}\, ,
\end{align}
\endgroup
for $X=$ BBH, BNS. Here, we have used that the BBH and BNS measurements are uncorrelated (meaning that the Fisher matrix can be written as a sum over the two populations) and that within one population, the number counts are uncorrelated with the distance and the mass, as shown in Sec.~\ref{sec:covariance}. The error on $v_0/c$ is then given by
\begin{align}
\sigma_{v_0/c}=\sqrt{\mathcal{F}^{-1}_{v_0/c}}\, .   
\end{align}

To compute $\sigma_{v_0/c}$, we need to know the values of the coefficients $\alpha^X_Y$ for the different populations and the different estimators. For the BBH, since threshold effects are negligible we have $\alpha^{\rm BBH}_N=\alpha^{\rm BBH}_{d_L}=\alpha^{\rm BBH}_{\cal{M}}=1$. For the BNS, we use the values calculated theoretically in Sec.~\ref{sec:alpha} and reported in Table~\ref{tab:BNSalpha}. 

The error, $\sigma_{v_0/c}$, is reported in Table~\ref{tab:error_known} for different cases. First, we compute the error from each estimator taken individually. We see that, as expected, the chirp mass estimator for BNSs gives the smallest uncertainty. The number count estimator for BNSs and for BBHs are the other two best ones. The number count for BNSs is very slightly better than for BBHs, due to threshold effects, that increase $\alpha_N$ from 1 to 1.08. Combining the three BBH estimators improves the constraints by 20\% compared to using the number counts alone, while combining the three BNS estimators improves the constraints by 32\%. Combining all estimators improves the constraints by 50\% compared to the number counts of BBHs alone (studied in~\citet{Mastrogiovanni:2022nya}). We also show the results for the top 3 estimators, i.e.\ mass estimator for BNSs and number count estimators for BBHs and BNSs. We see that the constraints are very similar to those obtained with all estimators.

The absolute error on the observer velocity is the same for the CMB and AGN case, however the relative error is reduced by a factor 5 for the AGN case, as can be seen from Table~\ref{tab:error_known}. Hence we see that a robust measurement of the observer velocity requires $10^6$ events if the observer velocity is consistent with the CMB dipole, but only $10^5$ events if the observer velocity is consistent with the AGN one, which is consistent with the detection efficiency results in Figs.~\ref{fig:BBHs_detectionprob} and~\ref{fig:BNSs_detectionprob}. This means in particular that if we do not detect a dipole with $10^5$ events, the GW dipole is in tension with the AGN dipole.

\subsection{Adding uncertainties on $\alpha$'s}
\label{sec:Fisher_erroralpha}

\begin{table*}
\centering
\caption{Fisher bound on the error $\sigma_{v_0/c}$, obtained from the combination of the six estimators assuming different uncertainties on the $\alpha$'s for the BNSs. In the top three rows, we consider a dipole consistent with the CMB one, while in the bottom three rows we consider a dipole consistent with the AGN one.}\label{tab:error_unknown}
\begin{tabular}{ccccc}
\toprule
& &10\% & 20\% & 50\% \\
\midrule
\parbox[t]{3mm}{\multirow{3}{*}{\rotatebox[origin=c]{90}{CMB}}}&$N_{\rm tot}=10^5$ & 110\%& 110\% & 112\% \\ 
&$N_{\rm tot}=10^6$ &35\% & 36\%& 40\% \\
&$N_{\rm tot}=10^7$ & 12\%& 13\%& 16\%\\
\midrule
\parbox[t]{3mm}{\multirow{3}{*}{\rotatebox[origin=c]{90}{AGN}}}&$N_{\rm tot}=10^5$ & 23\%& 24\%&   28\%\\ 
&$N_{\rm tot}=10^6$ & 8\%&  9\% &  11\% \\
&$N_{\rm tot}=10^7$ & 3\%& 3\% &  4\%\\
\bottomrule
\end{tabular}
\end{table*}

Whereas for the BBHs the $\alpha$ coefficients are known and equal to 1, for BNSs it is not the case. These coefficients are affected by threshold effects. As shown in Sec.~\ref{sec:alpha}, these coefficients can be computed, assuming a given model for the population of BNSs. The uncertainty on the model will directly impact the determination of the $\alpha$'s. To account for these uncertainties in the Fisher computation, we add the coefficients in the signal, and assign a corresponding error in the covariance matrix. We then compute how this uncertainty degrades the constraints on $v_0/c$. We consider the combination of all estimators and compute the error on $v_0/c$ for different values of the uncertainties on the $\alpha$'s. The signal is written as
\begin{align}
S\equiv
\left(
\hat{v}^{\rm BBH}_{N}, 
\hat{v}^{\rm BBH}_{d_L},
\hat{v}^{\rm BBH}_{\cal{M}},
\hat{v}^{\rm BNS}_{N}, 
\hat{v}^{\rm BNS}_{d_L},
\hat{v}^{\rm BNS}_{\cal{M}},
\alpha^{\rm BNS}_{N},
\alpha^{\rm BNS}_{d_L},
\alpha^{\rm BNS}_{\cal{M}}
\right)\, .  
\end{align}
The Fisher matrix contains now four parameters: $\theta_1=v_0/c, \theta_2=\alpha^{\rm BNS}_{N}, \theta_3=\alpha^{\rm BNS}_{d_L}$ and $\theta_4=\alpha^{\rm BNS}_{\cal{M}}$ and it is given by
\begin{align}
 \mathcal{F}_{ab}= \sum_{ij}\frac{\partial S_i}{\partial \theta_a} {\rm cov}(S_i,S_j)^{-1}   \frac{\partial S_j}{\partial \theta_b} \, .
\end{align}
The covariance matrix has the form
\begingroup
\renewcommand*{\arraystretch}{1.4}
\begin{align}
\label{eq:covalpha}
&{\rm cov}= \\   
&\begin{pmatrix}
{\rm cov}_{\rm BBH} & 0 & 0 & 0 & 0\\
0 & {\rm cov}_{\rm BNS} & 0 & 0 & 0\\
0 & 0 & \left(x\cdot\alpha^{\rm BNS}_{N}\right)^2& 0 & 0\\
0 &  0 & 0 &\left(x\cdot\alpha^{\rm BNS}_{d_L}\right)^2& 0 \\
0 &  0 & 0 & 0 &\left(x\cdot\alpha^{\rm BNS}_{\cal{M}}\right)^2 
\end{pmatrix}\nonumber
\end{align}
\endgroup
where the $3\times3$ blocks ${\rm cov}_{\rm BBH}$ and ${\rm cov}_{\rm BNS}$ are given by Eq.~\eqref{eq:covX}. In Eq.~\eqref{eq:covalpha}, $x$ denotes the relative uncertainty on the determination of the $\alpha$ parameters for the BNS population, that we assume here to be the same. We consider three cases for $x$: 10\%, 20\% and 50\%. The error on $v_0/c$ is then given by
\begin{align}
\sigma_{v_0/c}=\sqrt{(\mathcal{F}^{-1})_{v_0/c\,v_0/c} }\, .
\end{align}
The results are reported in Table~\ref{tab:error_unknown}. Comparing with the combination of all estimators (second last column) in Table~\ref{tab:error_known}, we see that having a 10\% uncertainty on all three $\alpha^{\rm BNS}$ degrades the constraints on $v_0/c$ by at most 1\% compared to the case where these parameters are assumed to be known perfectly. Increasing this uncertainty to 50\% degrades the constraints by at most 5\%. Hence, even if our modelling of the threshold effects is not very accurate, the degradation remains small.

Comparing Table~\ref{tab:error_unknown} with the BBH column of Table~\ref{tab:error_known}, we see that even in the case where the uncertainty on the $\alpha$ is as large as 50\%, we still gain information by including BNSs. For example, for $10^6$ events and a dipole consistent with the CMB one, we gain 29\% in the measurement of $v_0/c$ by adding BNSs. If we can model the $\alpha$'s with a precision of 20\%, this gain increases to 36\% (38\% for a 10\% precision). Seen as a function of the $\alpha^{\rm BNS}$ uncertainty, the constraints from all estimators are upper-bounded by the constraints from the combination of the three BBH estimators. At worst, if the uncertainty on the $\alpha^{\rm BNS}$ is too large, no information is gained by adding BNSs.

\section{Conclusions}
\label{sec:conclusion}

In this paper we have developed a robust framework to measure the cosmic dipole using GW detections. Contrary to radio sources or quasars, for which only the sky position can be used, GWs have the advantage of providing three quantities that are affected by the observer velocity: sky position, luminosity distance and redshifted chirp mass. We have developed estimators of these three dipoles, and we have calculated their variance and covariance. We have found that the mass and distance estimators are partially correlated, but that they are both uncorrelated with the number count estimator. Combining the three of them does therefore increase the detectability of the dipole.

BBHs have the advantage over BNSs to be unaffected by threshold effects, since all sources within the frequency range of ET and CE will have SNR above threshold. On the other hand, a significant fraction of BNSs will have an SNR below threshold, meaning that threshold effects are relevant in this case. The dipole from BNSs can of course be detected even without knowing the amplitude of threshold effects. However, to interpret the results and determine if the amplitude is consistent or not with the CMB dipole, it is necessary to have a modelling of these effects. We have developed such a modelling and computed the amplitude of threshold effects. For our population model, we have found that these effects are small, of the order of 10\% at most for all three estimators. The amplitudes of these effects depend of course on the population model that is used, however we expect that the order of magnitude we estimated will not change when changing the  details of the population model. This shows that it is worth including BNSs to measure the observer velocity and test the isotropy of the Universe. 

Comparing the three BNS and three BBH estimators, we have found that the BNS chirp mass estimator is the one with higher detectability, i.e.\ lower variance. This is due to the fact that the variance is fully dominated by shot noise, which generates fluctuations in the radial distribution of sources, consequently changing the mean mass per pixel. Since the intrinsic mass distribution of BNSs is very narrow, this shot noise contribution is mainly due to the redshift dependence in the chirp mass, which is significantly smaller than the spread in luminosity distance. After the BNS chirp mass, the other two best estimators are the number counts of BBHs and BNSs. 

Overall, we have found that combining all events, we need a few $10^6$ events to detect a dipole consistent with the CMB one. On the other hand, if the dipole is consistent with the AGN one, we should detect it with $10^5$ events. This can be achieved already after one year of observation. In this context, the fact that threshold effects are small is crucial, since it ensures that they cannot boost the dipole by a factor 5, thus mimicking the amplitude of the AGN dipole (which is 5 times larger than the CMB one). Hence, if we see results consistent with the AGN dipole, we can robustly conclude that it is not due to threshold effects, but rather to a large intrinsic anisotropy of the large scale structure.

\section*{Acknowledgements}
N.G. and C.B. acknowledge funding from the European Research Council (ERC) under the European Union’s Horizon 2020 research and innovation program (Grant agreement No.~863929; project title ``Testing the law of gravity with novel large-scale structure observables"). C.B. is also supported by the Swiss National Science Foundation. S.F. is supported by the Fonds National Suisse, grant $200020\_191957$, and by the SwissMap National Center for Competence in Research. The work of G.C. is supported by CNRS and G.C and M.P. acknowledge support from the Swiss National Science Foundation (Ambizione grant, ``Gravitational wave propagation in the clustered universe"). S.M. acknowledges the support of the computing facilities at INFN Rome of the Amaldi Research center funded by the MIUR program
“Dipartimento di Eccellenza” (CUP: B81I18001170001).

\section*{Data Availability}

The simulations and numerical code underlying this paper are available on GitHub \href{https://github.com/simone-mastrogiovanni/cosmic_dipole_GW_3G}{\faGithub}.

\bibliographystyle{mnras}
\bibliography{GWbib} 

\appendix

\section{Computation of the mean and variance}\label{app:variance}

\subsection{Mean of the luminosity distance estimator}

The mean of the luminosity distance estimator~\eqref{dlObservable} is given by
\begin{align}
\langle \hat{v}_{d_L-\bn'}\rangle= \frac{-3}{N_{\rm sky}}\sum_{i=1}^{N_{\rm sky}}\sum_{j=1}^{N^{\rm det}_{i}} \left\langle\frac{(d_L)^i_j}{\hat{d}_L N^{\rm det}_{i}}\right\rangle  (\bn_i\cdot\bn')\, . \label{eq:mean}
\end{align}
The number of detections in angular pixel $i$ is given by
\begin{align}
N^{\rm det}_{i}=\int {\rm d}r \frac{{\rm d}N_{\rm det}}{{\rm d}\Omega {\rm d}r}(r,\bn_i)\Delta\Omega +\Delta N^i\, ,    
\end{align}
where $\Delta N^i$ is the shot noise contribution. The luminosity distance can be written as
\begin{align}
 (d_L)^i_j=d_L(r_j,\bn_i)+(\Delta d_L)^i_j\, ,   
\end{align}
where $(\Delta d_L)^i_j$ denotes the error in the measurement of the distance. 
Assuming that the shot noise is small compared to the number of events in the angular pixel, and that the error in the distance measurement is small compared to the distance, we can expand each term in Eq.~\eqref{eq:mean} keeping only linear contributions in $\Delta N^i$ and $(\Delta d_L)^i_j$. Since the mean of these quantities is zero: $\langle\Delta N^i \rangle= \langle(\Delta d_L)^i_j \rangle=0$, Eq.~\eqref{eq:mean} becomes
\begin{align}
&\label{eq:comput_mean} \langle \hat{v}_{d_L-\bn'}\rangle=\\
& \frac{-3\sum_{i=1}^{N_{\rm sky}}\rbr{\Delta\Omega\int {\rm d}r \frac{{\rm d}N_{\rm det}}{{\rm d}\Omega {\rm d}r}(r,\bn_i)}^{-1}\sum_{j=1}^{N_{\rm det}^{i}} d_L(r_j,\bn_i)(\bn_i\cdot \bn')}
{\sum_{m=1}^{N_{\rm sky}}\rbr{\Delta\Omega\int {\rm d}r \frac{{\rm d}N_{\rm det}}{{\rm d}\Omega {\rm d}r}(r,\bn_m)}^{-1}\sum_{n=1}^{N_{\rm det}^{m}} d_L(r_n,\bn_m)} \, . \nn
\end{align}
To compute this, we divide each solid angle $i$ into $N_r$ bins in $r$ of size $\Delta r$, and we rewrite the sum over $j$ as
\begin{align}
 \sum_{j=1}^{N_{\rm det}^{i}} d_L(r_j,\bn_i)&=\sum_{k=1}^{N_r} \frac{{\rm d}N_{\rm det}}{{\rm d}\Omega {\rm d}r}(r_k,\bn_i)\Delta r \Delta \Omega\ d_L(r_k,\bn_i)\, ,
 \label{DistanceSum}
\end{align}
where $r_k$ denotes the center of the bin number $k$ in $r$.
In the continuous limit, i.e.\ taking $N_r\rightarrow\infty$ corresponding to $\Delta r\rightarrow 0$, this becomes
\begin{align}
\Delta\Omega\int {\rm d}r \frac{{\rm d}N_{\rm det}}{{\rm d}\Omega {\rm d}r}(r,\bn_i)d_L(r,\bn_i)  \, .  
\end{align}
Similarly, in the continuous limit the sum over $N_{\rm sky}\rightarrow \infty$ becomes
\begin{align}
 \sum_i^{N_{\rm sky}}F(r,\bn_i)=\frac{1}{\Delta\Omega}\int {\rm d}\Omega\, F(r,\bn)\, .   
\end{align}
With this, the numerator in Eq.~\eqref{eq:comput_mean} becomes
\begin{align}
&\frac{-3}{\Delta\Omega}\int {\rm d}\Omega\frac{\int {\rm d}r \frac{{\rm d}N_{\rm det}}{{\rm d}\Omega {\rm d}r}(r,\bn)d_L(r,\bn)}{\int {\rm d}r \frac{{\rm d}N_{\rm det}}{{\rm d}\Omega {\rm d}r}(r,\bn)} (\bn\cdot\bn') \nn\\
&= \frac{-3}{\Delta\Omega}\int {\rm d}\Omega\, d_L^{\rm mean}(\bn)(\bn\cdot\bn')\nn \\
&=N_{\rm sky}d_L^{(0)}\alpha_{d_L}\frac{v_0}{c}\cos\theta'\, ,
\end{align}
where we have used that $\Delta\Omega=4\pi/N_{\rm sky}$. Similarly, the denominator in Eq.~\eqref{eq:comput_mean} becomes 
\begin{align}
&\frac{1}{\Delta\Omega}\int {\rm d}\Omega\frac{\int {\rm d}r \frac{{\rm d}N_{\rm det}}{{\rm d}\Omega {\rm d}r}(r,\bn)d_L(r,\bn)}{\int {\rm d}r \frac{{\rm d}N_{\rm det}}{{\rm d}\Omega {\rm d}r}(r,\bn)} \nn\\
&= \frac{1}{\Delta\Omega}\int {\rm d}\Omega\, d_L^{\rm mean}(\bn)\nn \\
&=N_{\rm sky}d_L^{(0)}\, .
\end{align}
Therefore, we obtain for the mean
\begin{align}
\langle \hat{v}_{d_L-\bn'}\rangle= \alpha_{d_L}\frac{v_0}{c}\cos\theta'=v_{d_L-\bn'}\, .
\end{align}

\subsection{Variance of the luminosity distance estimator}

\begin{figure*}
    \centering
    \includegraphics[width=\textwidth]{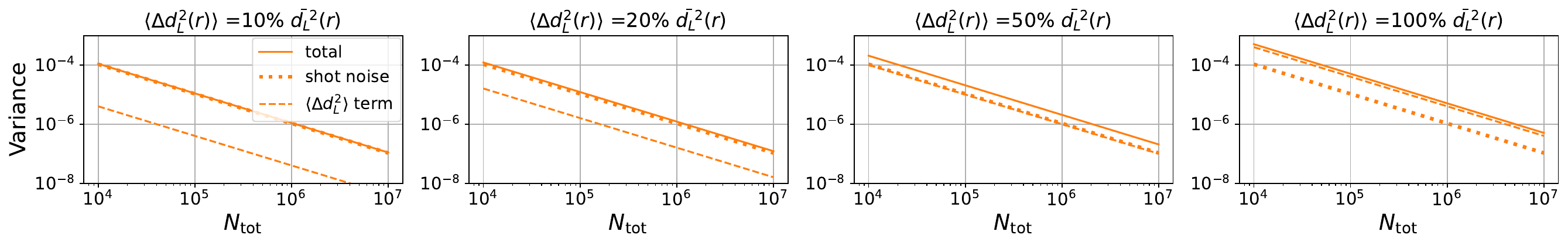}
    \caption{Variance of the luminosity distance estimator for BNSs, plotted as a function of the total number of events $N_{\rm tot}$. The different panels are for different values of the relative error on $d_L$ (assumed to be independent of redshift). We show separately the shot noise contribution and the distance-uncertainty contribution, as well as the total.}
    \label{fig:vardL_plots_BNS}
\end{figure*}

To compute the variance of the luminosity distance, we use Eq.~\eqref{eq:ratiovar}. We start by computing the variance of the variable $X$ defined in Eq.~\eqref{eq:X},
\begin{align}
&{\rm var}(X)=\langle X^2\rangle-\langle X\rangle^2\nn \label{eq:varianceX}\\
&=\frac{9}{ N_{\rm sky}^2}\sum_{i=1}^{N_{\rm sky}}\sum_{m=1}^{N_{\rm sky}}\sum_{j=1}^{N^i_{\rm det}}\sum_{n=1}^{N^m_{\rm det}}
(\bn_i\cdot\bn')(\bn_m\cdot\bn')\nn\\
&\qquad\times\left(\left\langle\frac{(d_L)^i_j(d_L)^m_n}{N^i_{\rm det}N^m_{\rm det}} \right\rangle-
\left\langle\frac{(d_L)^i_j}{N^i_{\rm det}} \right\rangle\left\langle\frac{(d_L)^m_n}{N^m_{\rm det}} \right\rangle\right)\,.
\end{align}
As done for the mean, we divide each solid angle $i$ into $N_r$ bins in $r$ of size $\Delta r$, and we rewrite the sum over $j$ as a sum over the $r$-bins. We obtain 
\begin{align}
&{\rm var}(X)=\frac{9}{ N_{\rm sky}^2}\sum_{i=1}^{N_{\rm sky}}\sum_{m=1}^{N_{\rm sky}}\sum_{k=1}^{N_r}\sum_{\ell=1}^{N_r}
(\bn_i\cdot\bn')(\bn_m\cdot\bn') \label{eq:var}\\
&\times\left(\left\langle\frac{N_p^{ik}(d_L)^i_k N_p^{m\ell}(d_L)^m_\ell}{N^i_{\rm det}N^m_{\rm det}} \right\rangle-
\left\langle\frac{N_p^{ik}(d_L)^i_k}{N^i_{\rm det}} \right\rangle\left\langle\frac{N_p^{m\ell}(d_L)^m_\ell}{N^m_{\rm det}} \right\rangle\right)\, ,\nonumber
\end{align}
where $N_p^{ik}$ denotes the number of events in the pixel in direction $\bn_i$ and centered at distance $r_k$ and $(d_L)^i_k=d_L(\bn_i,r_k)$. Neglecting the dipole contribution, which does not contribute to the variance, we have
\begin{align}
N_p^{ik}&=\bar{N}_p^k+\Delta N_p^{ik}\,,\\
N^i_{\rm det}&=\bar{N}_{\rm det}+\sum_j \Delta N_p^{ij}\,,\\
(d_L)^i_k&=(\bar{d}_L)_k +(\Delta d_L)^i_k\,.
\end{align}
Here $(\Delta d_L)^i_k$ is the error for one measurement of $d_L$ in angular bin $i$ and radial bin $k$. Inserting this into Eq.~\eqref{eq:var}, keeping only the terms quadratic in $\Delta N_p^{ik}$ and $(\Delta d_L)^i_k$, and using that the two types of error are uncorrelated we obtain
\begin{align}
&{\rm var}(X)=\nn\\
&\simeq\frac{9}{ N_{\rm sky}^2}\sum_{i=1}^{N_{\rm sky}}\sum_{m=1}^{N_{\rm sky}}\sum_{k=1}^{N_r}\sum_{\ell=1}^{N_r}(\bn_i\cdot\bn')(\bn_m\cdot\bn')\frac{\bar{N}_p^{k}\bar{N}_p^{\ell}(\bar{d}_L)_k (\bar{d}_L)_\ell}{\bar{N}_{\rm det}^2} \label{eq:var2}
 \nn\\
&\times\Bigg(\left\langle \frac{\Delta N_p^{ik}}{\bar{N}_p^{k}}\frac{\Delta N_p^{m\ell}}{\bar{N}_p^{\ell}} \right\rangle-\sum_n\left\langle \frac{\Delta N_p^{ik}}{\bar{N}_p^{k}}\frac{\Delta N_p^{m n}}{\bar{N}_{\rm det}} \right\rangle
-\sum_j\left\langle \frac{\Delta N_p^{m\ell}}{\bar{N}_p^{\ell}}\frac{\Delta N_p^{ij}}{\bar{N}_{\rm det}} \right\rangle\nonumber\\
&+\sum_{j,n=1}^{N_r}\left\langle \frac{\Delta N_p^{ij}}{\bar{N}_{\rm det}}\frac{\Delta N_p^{m n}}{\bar{N}_{\rm det}} \right\rangle+ \left\langle\frac{(\Delta d_L)^i_k(\Delta d_L)^m_\ell}{(\bar{d}_L)_k(\bar{d}_L)_\ell} \right\rangle\Bigg)\, .
\end{align}
We then use that
\begin{align}
& \langle \Delta N_p^{ik} \Delta N_p^{m\ell}\rangle=\bar{N}_p^k \delta_{im}\delta_{k\ell}\,.  \label{eq:shotnoise}
\end{align}
With this, the first 4 terms in the parenthesis of Eq.~\eqref{eq:var2} become
\begin{align}
 \frac{1}{\bar{N}_p^k}\delta_{im}\delta_{k\ell}-\frac{1}{\bar{N}_{\rm det}}\delta_{im}\, .
\end{align}
To compute the error from the luminosity distance we need to rewrite the sum over r-bins as a sum over events:
\begin{align}
\sum_{k=1}^{N_r}\sum_{\ell=1}^{N_r}\bar{N}_p^k\bar{N}_p^\ell  \langle (\Delta d_L)^{i}_k (\Delta d_L)^{m}_\ell\rangle=\sum_{j=1}^{N^i_{\rm det}}\sum_{n=1}^{N^m_{\rm det}} \langle (\Delta d_L)^{i}_j (\Delta d_L)^{m}_n\rangle\, , \label{eq:error_dL}
\end{align}
and we use that the errors for different events are uncorrelated
\begin{align}
\langle (\Delta d_L)^{i}_j (\Delta d_L)^{m}_n\rangle=\langle (\Delta d_L(r_j))^2  \rangle \delta_{im}\delta_{jn}\, .
\end{align}
Inserting this in Eq.~\eqref{eq:error_dL} we obtain
\begin{align}
\sum_{k=1}^{N_r}\sum_{\ell=1}^{N_r}\bar{N}_p^k\bar{N}_p^\ell  \langle (\Delta d_L)^{i}_k (\Delta d_L)^{m}_\ell\rangle&=\sum_{j=1}^{N^i_{\rm det}}\langle (\Delta d_L(r_j))^2  \rangle \delta_{im}\\
&=\sum_{k=1}^{N_r}\bar{N}_p^k \langle (\Delta d_L(r_k))^2  \rangle \delta_{im}\nonumber\, .
\end{align}
Inserting this in the variance and using that $N_{\rm sky}\bar{N}_{\rm det}=N_{\rm tot}$  we obtain 
\begin{align}
{\rm var}(X)=&\frac{3 N_{\rm sky}}{ N_{\rm tot}^2}\Bigg\{\sum_{k=1}^{N_r} {\bar{N}_p^k}\langle(\Delta d_L)^2_k \rangle \label{eq:varX}\\
&\sum_{k=1}^{N_r}(\bar{d}_L)_k^2 \bar{N}_p^k-\frac{1}{\bar{N}_{\rm det}}\left(\sum_{k=1}^{N_r}(\bar{d}_L)_k\bar{N}_p^k\right)^2 \Bigg\}\,, \nonumber
\end{align}
where we have used that
\begin{align}
\sum_{i=1}^{N_{\rm sky} }(\bn_i\cdot\bn')^2=\frac{1}{\Delta\Omega}\int {\rm d}\Omega\, (\bn\cdot\bn')^2=\frac{N_{\rm sky}}{3}\label{eq:variance_angle}\,.
\end{align}

The calculation of the variance of $Y=\hat{d}_L$ is exactly the same, except for the factor $-3$ which is not in $Y$ and the fact that $\hat{d}_L$ does not contain the product $\bn_i\cdot\bn'$ that is present in $X$, see Eq.~\eqref{eq:X}. As a consequence, Eq.~\eqref{eq:variance_angle} is replaced by
\begin{align}
\sum_{i=1}^{N_{\rm sky} }=N_{\rm sky}\, ,
\end{align}
leading to ${\rm var}(Y)=3{\rm var}(X)$. 
 
The calculation of the covariance between $X$ and $Y$ is also similar. In this case, Eq.~\eqref{eq:variance_angle} is replaced by
\begin{align}
\sum_{i=1}^{N_{\rm sky} }\bn_i\cdot\bn'=\frac{1}{\Delta\Omega}\int {\rm d}\Omega\, \bn\cdot\bn'=0\,,
\end{align}
meaning that the covariance is exactly zero.

We can easily show that the mean $X$ is proportional to $d_L^{(0)}\cdot v_0/c$, whereas the mean of $Y$ is equal to $d_L^{(0)}$. As a consequence, the second term in Eq.~\eqref{eq:ratiovar} is suppressed by a factor $(v_0/c)^2$ with respect to the first one and we can neglect it. We therefore have
\begin{align}
{\rm var}\left(\hat{v}_{d_L-\bn'} \right)=\frac{{\rm var}(X)}{\left(d_L^{(0)} \right)^2}\, .
\end{align}

In Fig.~\ref{fig:vardL_plots_BNS}, we show the variance of the luminosity distance for BNSs, as a function of $N_{\rm tot}$ and for different values of the measurement uncertainty on $d_L$. Fig.~\ref{fig:vardL_plots} shows the same for BBHs.

\section{Derivation in redshift space}\label{app:intz}

In Eqs.~\eqref{alternativedL}, \eqref{dipoleM} and \eqref{CountDipole}, we express luminosity distance, chirp mass and number count dipoles in terms of integrals over $r$-dependent quantities. The comoving distance $r$ is however not an observable quantity. On the other hand, the redshift $z$ is observable. In this Appendix we first show how the dipoles at fixed $z$ differ from the dipoles at fixed $r$. We then demonstrate that once we integrate over all sources, we obtain the same result, as expected.

We sketch the process as follows: consider any function $f(x, \epsilon)$ depending on one variable $x$ which should be changed, and a small parameter $\epsilon$ ($f$ may have other parameters that are neither small nor involved in the change of variable), such that expanding to first order we may write $f(x, \epsilon) =  f^{(0)}(x)+\epsilon f^{(1)}(x)$. Assume further that $f$ vanishes as $x\to 0, x\to\infty$.
The goal is to perform the $\epsilon$-dependent change variable from $x$ to $\Tilde{x} = x + \epsilon g(x)$, i.e. $x = \Tilde{x} - \epsilon g(\Tilde{x})$. We obtain
\begin{equation}
    f(x, \epsilon) = f\big(\Tilde{x}-\epsilon g(\Tilde{x}), \epsilon\big) \simeq f^{(0)}(\tilde{x})+\epsilon \bigg(f^{(1)}(\tilde{x}) -\frac{{\rm d}f^{(0)}(\Tilde{x})}{{\rm d} \tilde{x}} g(\Tilde{x})\bigg)\,. 
    \label{DipoleTilde}
\end{equation}
Thus, in the change of variable, the same function $f$ gains one extra term at first order.

In our case, $x$ can be thought to be $\bar{z}$ (since a fixed value of $r$ corresponds to a fixed value of the background redshift $\bar{z}$), and $\Tilde{x} \equiv z \simeq \bar{z}- (1+\bar{z})\,\bnv$. Hence we have: $\epsilon \equiv \bnv$ and $g(x) = -(1+x)$.

With this knowledge, we can write the number count density of detected sources per redshift bin.
Recalling that
\begin{equation}
    r(\bar{z}) = c \int_0^{\bar{z}} \frac{{\rm d}z'}{H(z')} = r\bigg(z+(1+z)\bnv\bigg)\,,
    \label{ComovingDistance}
\end{equation}
we use Eq.~\eqref{DipoleTilde} with \begin{align}
    f &\equiv \frac{{\rm d}N_{\rm det}}{{\rm d}\Omega {\rm d}\bar{z} {\rm d}m_{1,2}} (\bar{z}, \bn, \mm, \rho>\rho_*)\nonumber \\&= \frac{c}{H(\bar{z})}  \frac{{\rm d}N_{\rm det}}{{\rm d}\Omega {\rm d}r {\rm d}m_{1,2}}(r(\bar{z}), \bn, \mm, \rho>\rho_*)\,,
\end{align} 
of which the functional form has been worked out in Eq.~\eqref{DipoleCountDensity} until the dipolar order. Here $H$ is the Hubble parameter.
We end up with 
\begin{align}
    &\frac{{\rm d}N_{\rm det}}{{\rm d}\Omega {\rm d}z {\rm d}m_{1,2}} (z, \bn, \mm, \rho>\rho_*) \nonumber \\=& \frac{{\rm d}\bar{z}}{{\rm d}z} \frac{{\rm d}N_{\rm det}}{{\rm d}\Omega {\rm d}\bar{z} {\rm d}m_{1,2}} \bigg(z+(1+z)\bnv, \bn, \mm, \rho>\rho_*\bigg)\,,
\end{align}
from which Eq.~\eqref{DipoleTilde} allows us to write the monopole as
\begin{align}
    &\frac{{\rm d}\bar{N}_{\rm det}}{{\rm d}\Omega {\rm d}z {\rm d}m_{1,2}} (z, \mm, \rho>\rho_*) = \nonumber \\ &\frac{c}{H(z)} \frac{{\rm d}\bar{N}_{\rm det}}{{\rm d}\Omega {\rm d}r {\rm d}m_{1,2}}\rbr{c \int_0^{z} \!\!\!\frac{{\rm d}z'}{H(z')}, \mm, \rho>\rho_*}\,,
\end{align}
while the dipole term is rewritten in terms of the monopole as
\begin{align}
   \bnv\Bigg\{ \frac{{\rm d}\bar{N}_{\rm det}}{{\rm d}\Omega {\rm d}z {\rm d}m_{1,2}} \left[3+s\left(\frac{1}{3}+ {\calA}\right)\right]+(1+z)\frac{{\rm d}}{{\rm d} z}\bigg(\frac{{\rm d}\bar{N}_{\rm det}}{{\rm d}\Omega {\rm d}z {\rm d}m_{1,2}}\bigg)\Bigg\}\,,
\end{align}
where $s$ and $\calA$ are the functions defined in Eqs.~\eqref{newdefhats} and~\eqref{newdefhatA}, evaluated at the arguments $(c \int_0^{z} \!\!\!\frac{{\rm d}z'}{H(z')}, \mm, \rho_*)$ and $(c \int_0^{z} \!\!\!\frac{{\rm d}z'}{H(z')}, \mm)$ respectively.

Once equipped with the dipole expansion of detected events number count density per redshift, we can proceed analogously to Sec.~\ref{sec:model} and introduce the dipoles in the luminosity distance and chirp mass distributions of detected events.

For the luminosity distance, the expansion in velocity at fixed $z$ has been done in ~\cite{Bonvin:2005ps} and it reads
\begin{equation}
    d_L(z,\bn)=\bar d_L(z)\rbr{1+\frac{\bn\cdot \bv_0}{\mathcal H(z)\bar r(z)}}\,,
\end{equation}
with $\mathcal H(z)$ the comoving Hubble parameter, and where $\bar r(z)$ is the monopole of the velocity expansion of $r$ at fixed $z$, which from the third expression in Eq.~\eqref{ComovingDistance} reads: $\bar{r}(z) = c \int_0^{z} \frac{{\rm d}z'}{H(z')}$.

The redshifted chirp mass does not have a dipole with respect to fixed observed redshift $z$ slices, since it is simply the product of the source chirp mass with $(1+z)$, which is obviously constant on a slice of constant $z$.

As expected, the dipoles at fixed $r$ are different from the dipoles at fixed $z$. Let us now see what happens once we integrate over $r$ and $z$.

Integrating Eq.~\eqref{DipoleTilde} between 0 and $\infty$, we obtain
\begin{align}
&\int\!\!\!{\rm d}x f^{(0)}(x) +\epsilon \int\!\!\!{\rm d}x f^{(1)}(x) = \int\!\!\!{\rm d}x f(x, \epsilon) = \int\!\!\!{\rm d}\Tilde{x} f\big(x(\Tilde{x}, \epsilon), \epsilon\big)\frac{{\rm d}x}{{\rm d}\Tilde{x}}\nonumber \\
 &=  \int\!\!\!{\rm d}\Tilde{x} f^{(0)}(\Tilde{x}) + \epsilon \int\!\!\!{\rm d}\Tilde{x} \bigg( f^{(1)}(\Tilde{x})-\frac{{\rm d}f^{(0)}(\Tilde{x})}{{\rm d} \tilde{x}} g(\Tilde{x})-f^{(0)}(\Tilde{x})\frac{{\rm d}g(\Tilde{x})}{{\rm d} \tilde{x}}\bigg)\nonumber \\
 &=  \int\!\!\!{\rm d}\Tilde{x} f^{(0)}(\Tilde{x}) + \epsilon \int\!\!\!{\rm d}\Tilde{x} f^{(1)}(\Tilde{x})\,,
\end{align} 
where in the last step we used integration by parts and the assumed asymptotic behaviour for $f$ to make the boundary term vanish. The equivalence of the integrals holds both for the $\epsilon^0$ and $\epsilon^1$ terms.

Note that if the integration boundaries over $x$ are $[0, \infty[$, they should in principle become $[\epsilon g(0), \infty[$ for the $\Tilde{x}$ integration. However, at first order in $\epsilon$, we may take integrals over $[0, \infty[$ for $\Tilde{x}$ as well, the correction between the two being of order $\epsilon^2$.

With this, we can build the monopoles and dipoles in the distribution of detected luminosity distances and chirp masses, analogously to Eqs.~\eqref{alternativedL} and \eqref{dipoleM}. The redshift integrals over the dipoles at fixed $z$ are equivalent to the $r$ integrals over dipoles at fixed $r$, by the above argument.

\bsp	
\label{lastpage}
\end{document}